\documentclass[aps,prd,nofootinbib,twocolumn,superscriptaddress,preprintnumbers,letterpaper,natbib,nofootinbib]{revtex4-1}

\pdfoutput=1
\usepackage{amssymb,amsmath,latexsym,mathrsfs}
\usepackage[normalem]{ulem}
\usepackage{url}
\usepackage{epsfig,graphicx,verbatim,xspace,multirow,mathtools}
\usepackage{enumitem}
\usepackage{graphicx}
\usepackage[usenames,dvipsnames]{color}
\usepackage[breaklinks,colorlinks,urlcolor=blue,citecolor=blue,linkcolor=magenta]{hyperref}
\usepackage{comment}
\usepackage{bm}
\usepackage[table]{xcolor}

\usepackage{hhline}
\usepackage{multirow}

\newcommand{\nint}{n_{\mathrm{int}}}
\newcommand{\zint}{z_{\mathrm{int}}}

\newcommand{\LCDM}{$\Lambda$CDM}
\newcommand{\eV}{\mathrm{eV}}

\makeatletter

\begin{document}

\title{Global view of neutrino interactions in cosmology:\\ The freestreaming window as seen by Planck}

\preprint{TUM-HEP-1406/22}
\author{Petter Taule}
\email{petter.taule@tum.de}
\thanks{ORCID: \href{https://orcid.org/0000-0002-6658-901X}{0000-0002-6658-901X}}
\affiliation{Physik-Department, Technische Universit{\"{a}}t, M{\"{u}}nchen, James-Franck-Stra{\ss}e, 85748 Garching, Germany}
\author{Miguel~Escudero}
\email{miguel.escudero@tum.de}
\thanks{ORCID: \href{https://orcid.org/0000-0002-4487-8742}{0000-0002-4487-8742}}
\affiliation{Physik-Department, Technische Universit{\"{a}}t, M{\"{u}}nchen, James-Franck-Stra{\ss}e, 85748 Garching, Germany}
\author{Mathias Garny}
\email{mathias.garny@tum.de}
\thanks{ORCID: \href{https://orcid.org/0000-0003-4056-6802}{0000-0003-4056-6802}}
\affiliation{Physik-Department, Technische Universit{\"{a}}t, M{\"{u}}nchen, James-Franck-Stra{\ss}e, 85748 Garching, Germany}

\begin{abstract}
Neutrinos are expected to freestream (i.e.\ not interact with anything) since they decouple in the early Universe at a temperature $T\sim 2\,{\rm MeV}$. However, there are many relevant particle physics scenarios that can make neutrinos interact at $T< 2\,{\rm MeV}$. In this work, we take a global perspective and aim to identify the temperature range in which neutrinos can interact given current cosmological observations. We consider a generic set of rates parametrizing neutrino interactions and by performing a full Planck cosmic microwave background (CMB) analysis we find that neutrinos cannot interact significantly for redshifts $2000 \lesssim z \lesssim 10^5$, which we refer to as the \emph{freestreaming window}. We also derive a redshift dependent upper bound on a suitably defined interaction rate $\Gamma_\text{nfs}(z)$, finding $\Gamma_\text{nfs}(z)/H(z)\lesssim 1-10$ within the freestreaming window. We show that these results are largely model independent under some broad assumptions, and contextualize them in terms of neutrino decays, neutrino self-interactions, neutrino annihilations, and majoron models. We provide examples of how to use our model independent approach to obtain bounds in specific scenarios, and demonstrate agreement with existing results. We also investigate the reach of upcoming cosmological data  finding that CMB Stage-IV experiments can improve the bound on $\Gamma_\text{nfs}(z)/H(z)$ by up to a factor $10$. Moreover, we comment on large-scale structure observations, finding that the ongoing DESI survey has the potential to probe uncharted regions of parameter space of interacting neutrinos. Finally, we point out a peculiar scenario that has so far not been considered, and for which relatively large interactions around recombination are still allowed by Planck data due to some degeneracy with $n_s$, $A_s$ and $H_0$. This scenario can be fully tested with CMB-S4.
\end{abstract}

\maketitle
{
   \hypersetup{linkcolor=black}
   \setlength\parskip{-0.2pt}
   \setlength\parindent{0.0pt}
}

\section{Introduction}\label{sec:intro}
Neutrinos are ubiquitous in cosmology and they represent a relevant component of the energy density of the Universe across its entire history. For example, after electron-positron annihilation and while the Universe is radiation dominated neutrinos represent around $40\%$ of the energy density of the Universe. This in turn means that we can use cosmological observations to shed light on the properties of the most elusive particles in the Standard Model~\cite{Dolgov:2002wy,Lesgourgues:2013sjj}. Prime examples of this power are the bounds that can be derived on the neutrino mass, $m_\nu$, and the number of effective relativistic neutrino species, $N_{\rm eff}$, from cosmic microwave background (CMB) observations~\cite{Planck:2018vyg}. Importantly, cosmological data can be used to test other relevant properties of neutrinos and in this work we will focus our attention on neutrino interactions.

\begin{figure*}[t]
\centering
		\includegraphics[width=0.9\textwidth]{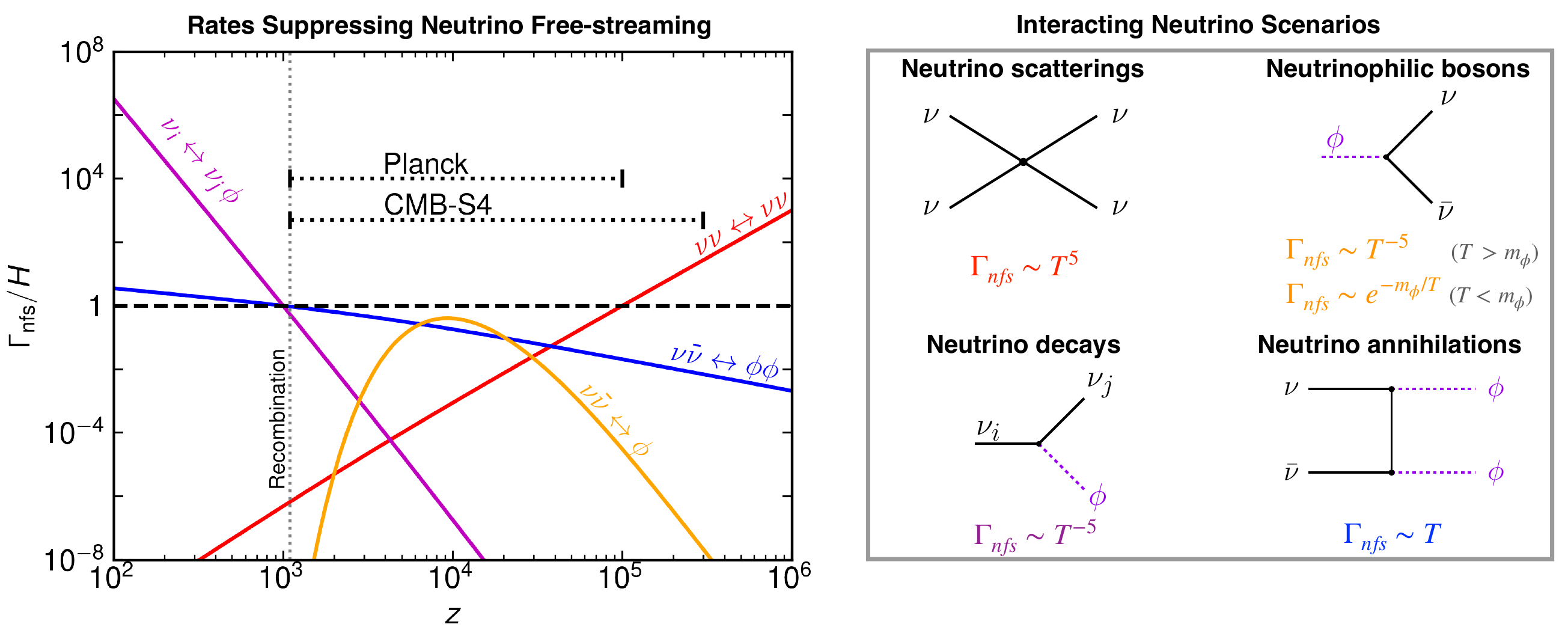}

\caption{\textit{Left panel:} Damping rates of the neutrino anisotropic stress ($\Gamma_{\rm nfs}$) as a function of redshift for various models of interacting neutrinos -- see Section~\ref{sec:applications} for a discussion of the various models. In dotted we highlight the sensitivity of Planck and CMB-S4 to these rates (provided that $\Gamma_{\rm nfs}/H \gtrsim 1$, see Figs.~\ref{fig:envelopes} and~\ref{fig:CMBS4forecast} for the actual constraints). \textit{Right panel:} summary of models with interacting neutrinos including the overall scaling of the rate suppressing neutrino freestreaming in the early Universe. }
\label{fig:summary}
\end{figure*}

In the Standard Model, neutrinos decouple from the primordial plasma at a temperature $T\sim 2\,{\rm MeV}$~\cite{EscuderoAbenza:2020cmq,Escudero:2018mvt}, when the Universe was only $t\sim 0.1\,{\rm s}$ old, and they do not interact with anything else afterwards apart from gravity. Given that neutrinos are ultrarelativistic until very late times, $z_{\rm nr}  \simeq 200 \,m_\nu/0.1\,{\rm eV}$, and that they do not interact, neutrinos are said to \textit{freestream}. The freestreaming nature of neutrinos makes them a very special component at the level of cosmological perturbations and this has important implications for CMB observations~\cite{Bashinsky:2003tk}. The reason for this is twofold. Firstly, since neutrinos do not interact, they represent the only species capable of developing a sizeable anisotropic stress (which is gradually generated by small velocity perturbations sourced by the primordial perturbation spectra). Secondly, the neutrino anisotropic stress, via Einstein's equations, directly affects metric perturbations which are the source of the CMB anisotropies. Indeed, the neutrino anisotropic stress together with  ultrarelativistic neutrino velocities leads to phase shifts on the CMB spectra that cannot be mimicked by standard cosmological parameters given adiabatic perturbations~\cite{Bashinsky:2003tk}. 

Current CMB observations are in excellent agreement with the Standard Model picture of three freestreaming neutrinos, see~\cite{Planck:2015fie,Planck:2018vyg}. However, there are many scenarios beyond the Standard Model that can affect neutrino freestreaming and make neutrinos interact in the early Universe. In this context, the CMB has proven to be a laboratory to test potential interactions of neutrinos with themselves and with other light species within various particle physics motivated frameworks. In particular, many groups have studied the cosmological implications of neutrinos undergoing strong self-interactions~\cite{Cyr-Racine:2013jua,Oldengott:2014qra,Lancaster:2017ksf,Oldengott:2017fhy,Kreisch:2019yzn,Park:2019ibn,Das:2020xke,RoyChoudhury:2020dmd,Brinckmann:2020bcn,Kreisch:2022zxp} (as mediated by e.g.\ a MeV-scale neutrinophilic boson), of neutrinos annihilating into massless scalars~\cite{Beacom:2004yd,Hannestad:2004qu,Bell:2005dr,Archidiacono:2013dua,Forastieri:2015paa,Forastieri:2019cuf,Venzor:2022hql}, of neutrinos interacting via decays and inverse decays with eV-scale neutrinophilic scalars~\cite{Chacko:2003dt,Escudero:2019gvw,Escudero:2021rfi,EscuderoAbenza:2020egd}, and of decaying neutrinos~\cite{Hannestad:2005ex,Basboll:2008fx,Escudero:2019gfk,Chacko:2019nej,Chacko:2020hmh,Barenboim:2020vrr,Chen:2022idm,Abellan:2021rfq}\footnote{Other somewhat related scenarios involve self-interacting sterile neutrinos~\cite{Hannestad:2013ana,Forastieri:2017oma,Song:2018zyl,Archidiacono:2020yey}, strongly interacting dark radiation~\cite{Baumann:2015rya,Brust:2017nmv,Blinov:2020hmc,Aloni:2021eaq}, or neutrinos interacting via long-range forces~\cite{Esteban:2021ozz,Esteban:2022rjk}. }.

At the cosmological level, the difference between all these models mainly resides in the temperature dependence of the interaction rate $\Gamma_{\rm nfs}$ parametrizing the damping of the neutrino anisotropic stress, to which we refer as \emph{non-freestreaming} rate and which plays a central role throughout our analysis. To highlight these differences we explicitly show the temperature evolution of this rate for various models with interacting neutrinos in Fig.~\ref{fig:summary}. For example, models where neutrinos annihilate into massless species affect neutrino freestreaming at low temperatures since $\Gamma_{\rm nfs}/H \sim T^{-1/2}$ (see blue line in Fig.~\ref{fig:summary}). Others, such as self-interacting neutrinos, suppress neutrino freestreaming at high temperatures since $\Gamma_{\rm nfs}/H \sim T^3$ (see red line in Fig.~\ref{fig:summary}). Furthermore, in scenarios with eV-scale neutrinophilic bosons the interaction is transient and affects neutrino freestreaming only within a certain window of redshifts (see orange line in Fig.~\ref{fig:summary}). 

While previous works have studied various aspects of neutrino freestreaming within particular frameworks we believe that there remain several aspects of this issue to be explored and we aim to address them in this study. Specifically, in this paper we aim to answer the following questions:

\begin{enumerate}
    \item \textit{What is the window of redshifts in which neutrinos need to freestream in order to be compatible with current CMB observations by Planck?} We believe that this is an important question that can globally be used by particle physics model builders to readily know the regions of interests within their models. We acknowledge that this question has been addressed within particular frameworks, see e.g.~\cite{Archidiacono:2013dua,Forastieri:2019cuf,Brinckmann:2020bcn,Chen:2022idm}, but we believe that it has not been addressed in a model-independent manner.
    \item \textit{Given this window, how efficient can the rates suppressing neutrino freestreaming be?} Once the neutrino freestreaming window is established, we would like to understand how large can the rate suppressing neutrino freestreaming be as compared to the expansion rate, $\Gamma_{\rm nfs}/H$. 
    \item \textit{What is the sensitivity of Stage-IV CMB experiments to neutrino freestreaming?} We would like to understand the global sensitivity of the next generation of CMB experiments to models with interacting neutrinos.
    \item \textit{What is the sensitivity of observations of the large-scale structure (LSS) to uncharted regions of parameter space with interacting neutrinos?} While the effect of neutrino interactions is most prominent in CMB observations, see e.g.~\cite{Hannestad:2004qu}, it is unclear whether LSS data can test regions of parameter space that are not already constrained by CMB data. 
\end{enumerate}

 With the purpose of answering these questions in mind, the structure of this work is as follows. In Section~\ref{sec:methodology} we describe the methodology we employ to account for the damping of neutrino freestreaming in a way that can directly be mapped to several relevant models of new physics. We also comment on the assumptions and limitations of our approach. Our main results are presented in Section~\ref{sec:results_models}, based on a full Planck legacy data analysis. There we show the existence of a redshift window in which neutrinos cannot interact given Planck data, and discuss its model (in-)dependence. In Section~\ref{sec:CMBS4} we explore the sensitivity of the future CMB-S4 experiment and in Section~\ref{sec:LSS} the potential reach of experiments probing the matter power spectrum to non-freestreaming neutrinos.  We show examples of how our generic constraints can be translated into bounds on new physics parameters within specific models with interacting neutrinos in Section~\ref{sec:applications}, where we also compare with existing literature. Finally, we present our conclusions in Section~\ref{sec:conclusions}. In addition, the interested reader can find several checks and details in Appendices~\ref{sec:approximations} and~\ref{sec:nint3appendix}.

\section{Damping of Neutrino Freestreaming and CMB analysis}\label{sec:methodology}

\subsection{Modeling of neutrino interactions: \\ Key ingredients and approximations}

Einsteins equations couple the neutrino anisotropic stress to the evolution of metric perturbations. These metric perturbations, in turn, are the source of the photon anisotropies which then form the basis of the CMB. Schematically, we have 
\begin{align}
    \sigma_\nu \to  \delta G_{\mu\nu} \to \delta T_{\mu \nu}|_\gamma  \to \delta T_\gamma\,,
\end{align}
where $\delta$ means a perturbation, $T_{\mu\nu}$ represents the stress energy tensor of a given species, $\sigma_\nu$ is the neutrino anisotropic stress which is intimately related to the traceless part of $\delta T^i_j|_\nu$, $G_{\mu \nu}$ is Einstein's tensor and $T_\gamma$ the CMB temperature.

In this work we are interested in scenarios where the neutrino anisotropic stress is damped by some interactions (i.e.\ $\sigma_\nu \to 0$). In order to model the growth, damping and potentially subsequent regeneration of anisotropic stress we shall model the evolution of the small neutrino perturbations using the synchronous gauge. Following standard methods~\cite{Ma:1995ey} we can describe the perturbations of massive neutrino perturbations as
\begin{subequations}
\label{eq:Boltzmannhierarchy}
\begin{align}
     \frac{d\Psi_0}{d\tau} &= -{qk\over \epsilon}\Psi_1
  	  +{1\over 6}\dot{h} {d\ln f_0\over d\ln q}
		\,, \\
     \frac{d\Psi_1}{d\tau}  &= {qk\over 3\epsilon} \left(\Psi_0
		      - 2 \Psi_2 \right) \,, \\
     \frac{d\Psi_2}{d\tau}  &= {qk\over 5\epsilon} \left(
	2\Psi_1 - 3\Psi_3 \right)
	 - \left( {1\over15}\dot{h} + {2\over5} \dot{\eta} \right)
	{d\ln f_0\over d\ln q} \,\nonumber \\
	&- a \,\Gamma_{\rm nfs} \,\Psi_2\,,   \label{eq:psi2}\\
    \frac{d\Psi_l}{d\tau} &= {qk \over (2l+1)\epsilon} \left[ l\Psi_{l-1}
        - (l+1)\Psi_{l+1} \right]\,\nonumber \\
	&- a \,\Gamma_{\rm nfs}\, \Psi_l   \,, \quad l \geq 3 \,. \label{eq:psil}
\end{align}
\end{subequations}
Here $\Psi_\ell(k,q,\tau)$ represents the contribution of the $\ell$th Legendre polynomial to the perturbed distribution function of neutrinos in Fourier space, i.e. $\delta f \simeq \Psi$. $f_0$ is the isotropic and homogeneous neutrino distribution function which we take to be a frozen Fermi-Dirac distribution with $T_\nu= T_\gamma/1.39578$ as expected from neutrino decoupling: $f_0(q) = [1+\exp(q/T_\nu^0)]^{-1}$, where $T_\nu^0$ is the neutrino temperature today. Here, $q$ represents the comoving momentum, $\epsilon$ is the comoving energy, $k$ represents a given comoving wave number, $\tau$ is the comoving time, $a$ is the scale factor, and $h$ and $\eta$ are the metric perturbations\footnote{We have disregarded tensor perturbations because their impact on the CMB anisotropies is small. We note, however, that the neutrino anisotropic stress energy tensor is in fact relevant in their evolution~\cite{Weinberg:2003ur}. However, we have explicitly checked that the effect of the neutrino interactions considered in this work do not lead to any relevant impact on the unlensed BB power spectrum.}. Finally, $\Gamma_{\rm nfs}$ is the rate at which neutrino freestreaming is suppressed. Before we turn to specific forms of this rate we would like to explicitly state various assumptions and approximations used to derive Eqs.~\eqref{eq:Boltzmannhierarchy}: 
\begin{enumerate}

    \item We use a relaxation time approximation, assuming that the neutrino interaction rate, $\Gamma_{\rm nfs}$, is independent of the momentum and the Fourier wave vector and only depends on temperature. Therefore, $\Gamma_{\rm nfs}$ corresponds to the average rate at which neutrino freestreaming is damped. This is a good approximation because in practice although our formalism tracks separately each neutrino $q$ mode, the cosmological neutrino mass bounds (even in the presence of interactions) restrict neutrinos to have small masses, $\sum m_{\nu} \lesssim 0.2\,{\rm eV}$. We shall see that neutrino interactions can only have relevant implications if they are active at $z \gtrsim 1000$, which means that neutrinos were ultrarelativistic at the time at which the interactions are active. This thus reproduces the same approach as used in e.g.~\cite{Cyr-Racine:2013jua,Forastieri:2019cuf,Escudero:2019gfk} but now allowing for a non-negligible neutrino mass. For a discussion of the effects of including the momentum dependence of the interactions, see~\cite{Oldengott:2014qra,Chen:2022idm}. 

    \item We have assumed that neutrinos form a single, ultrarelativistic fluid with the energy density expected in the Standard Model. This assumption is broadly justified in all scenarios of interacting neutrinos in cosmology presently in the literature (as highlighted in Fig.~\ref{fig:summary}). For example, in the case of neutrino scatterings there are indeed no new light BSM states and the approximation applies exactly. In the case of neutrino annihilations into light states, since these states are necessarily lighter than neutrinos this means that they are also relativistic. This happens similarly in the case of neutrino decays. In addition, the total energy density of the joint system is similar to the SM value until $T_\nu \lesssim m_\nu$ as a result of energy density conservation. At such low temperatures neutrinos cannot affect the primary CMB anisotropies and thus the effect of this is small. In addition, Refs.~\cite{Hannestad:2004qu} and~\cite{Barenboim:2020vrr} explicitly confirm the small change in the total energy density of the total fluid of neutrinos and BSM states in such scenarios. Finally, in the case of light bosons decaying into neutrinos the situation is a little bit different because these states need to be heavier than neutrinos and thus can easily become non-relativistic before recombination. However, it has been shown that in most cases the contribution to the energy density of the massive boson to the joint neutrino+boson system is typically small, $\lesssim O(10)\%$~\cite{EscuderoAbenza:2020cmq}. Furthermore, in such scenario the effect of reducing the neutrino freestreaming is much more significant than the modified expansion history provided that the rate of interactions is substantially large $\Gamma_{\rm nfs}/H \gg 1$, see~\cite{Escudero:2019gvw}. In order to deal with the region of $\Gamma_{\rm nfs}/H \sim 1$ one would need to take into account the effect of the mass of the boson in the sound speed and equation of state of the system. This, however, is an effect that would only alter a small region of parameter space under our study and we neglect it. Thus, in general this approximation is indeed well met in most of the models already present in the literature and within the testable parameter space. 
 \end{enumerate}

Finally, we refer the reader to Section~\ref{sec:applications} where we demonstrate that taking these approximations one can recover results that have been obtained in specific neutrino interacting scenarios in the literature.

\subsection{Rates suppressing neutrino freestreaming}
The power of the approximations described above is that they allow for a fast solution of the neutrino perturbations for many different scenarios suppressing neutrino freestreaming. Here, since we are aiming to find out in a model independent way the shape of the redshift window in which neutrino freestreaming is essential to explain CMB observations, we will consider an array of rates suppressing neutrino freestreaming. Firstly, we consider rates that scale as power-laws in temperature $\Gamma_{\rm nfs} \propto T^{n_{\rm int}}$, and that we explicitly describe by
\begin{align}\label{eq:Gamma_powerlaw}
    \Gamma_{\rm nfs}(z,z_{\rm int}) = H(z_{\rm int}) 
    \left(\frac{1+z}{1+\zint}\right)^{n_{\rm int}}\,.
\end{align}
Here $H(z_{\rm int})$ is the Hubble rate at the redshift $z_{\rm int}$ at which $\Gamma_{\rm nfs}(z_{\rm int}) = H(z_{\rm int})$, and $n_{\rm int}$ is a power-law index. For our analysis, we consider the  power-law indices:
\begin{align}
    n_{\rm int} &= [5,\,4,\,3]\qquad \qquad \,\,\text{(High-z interactions)}\,,\\
    n_{\rm int} &= [-5,\,-3,\,-1,1]\quad \text{(Low-z interactions)}\,.
\end{align}
We have split them into scenarios where the interactions become relevant at high redshift and then eventually turn off, and those that become relevant at low redshift. This distinction is easy to follow since $H\propto T^2$ in a radiation dominated Universe and $H\propto T^{3/2}$ in a matter dominated one. We note that these power-laws are phenomenological but  actually reproduce many well motivated scenarios. In particular, $n_{\rm int} = -5$ corresponds to neutrino decays~\cite{Barenboim:2020vrr,Chen:2022idm}, $n_{\rm int} = 1$ to neutrino annihilations into massless states~\cite{Oldengott:2014qra,Forastieri:2019cuf}, and $n_{\rm int} = 5$ to neutrino self-interactions~\cite{Cyr-Racine:2013jua,Lancaster:2017ksf}. Finally, we note that given that $H\propto T^2$ the rate with $n_{\rm int} = 3$ corresponds to a neutrino interaction that is almost constant over temperature which could lead to a qualitatively different behavior compared to the other cases.

In addition, we consider a family of rates of neutrino freestreaming that are transient in redshift as motivated by eV-scale neutrinophilic bosons~\cite{Chacko:2003dt,Escudero:2019gvw,Escudero:2021rfi,Barenboim:2020vrr,Chen:2022idm}. In particular, we describe the rate by the following function:
\begin{align}
    \Gamma_{\rm nfs}(z, \,z_\text{int}^\text{max},\,\Gamma/H|_{\rm nfs}^{\rm max}) &= \Gamma/H|_{\rm nfs}^{\rm max}\, \frac{H(z_\text{int}^\text{max})}{C} \\ 
    &\times K_2(a \, x)\, x^3 
    \left[   K_1 (ax)/K_2 (ax)    \right]^b\nonumber \,,
    \label{eq:bessel_int_improved}
\end{align}
where $x = (1+z_\text{int}^\text{max})/(1+z)$, and $z_\text{int}^\text{max}$ corresponds to the redshift at which the rate divided by the Hubble rate becomes maximal. The parameter $\Gamma/H|_{\rm nfs}^{\rm max}$ then controls the value of $\Gamma_{\rm nfs}/H$ at the maximum. We consider three different choices for the power-law index: $b = 0,\,2,\,4$, which determines the slope for $z\gg z_\text{int}^\text{max}$, where $\Gamma_{\rm nfs} \propto T^{-(1+b)}$. We set $a = 4.7$ which ensures that the rate in each of these scenarios becomes maximal at $z=z_\text{int}^\text{max}$. Finally, by imposing the normalization condition $\Gamma_{\rm nfs}/H|_{z_\text{int}^\text{max}} = \Gamma/H|_{\rm nfs}^{\rm max}$ we can numerically fix the constant $C$ for these cases. The values we consider are:
\begin{align}
    b &= 0,\,\,\,C =  1/130  \,,\\
    b &= 2,\,\,\,C =  1/240  \,,\\
    b &= 4,\,\,\,C =  1/420  \,.
\end{align}
Again, although phenomenological in nature, we note that these values are motivated from a particle physics viewpoint in the presence of decays and inverse decays of neutrinos (or particles interacting with them). In particular, $b= 0$ would correspond to the rate controlling the background evolution of such a system~\cite{Escudero:2019gvw}, $b=2$ represents the heuristic rate suppressing neutrino freestreaming assuming a random walk~\cite{Chacko:2003dt}, and $b =4$ represents the best motivated scenario and roughly matches the temperature dependence of the neutrino freestreaming suppression rate in scenarios with decays and inverse decays~\cite{Barenboim:2020vrr,Chen:2022idm}. For reference, this rate is shown in orange in Fig.~\ref{fig:summary}.

\begin{figure*}[t]
\centering
	\includegraphics[width=0.7\textwidth]{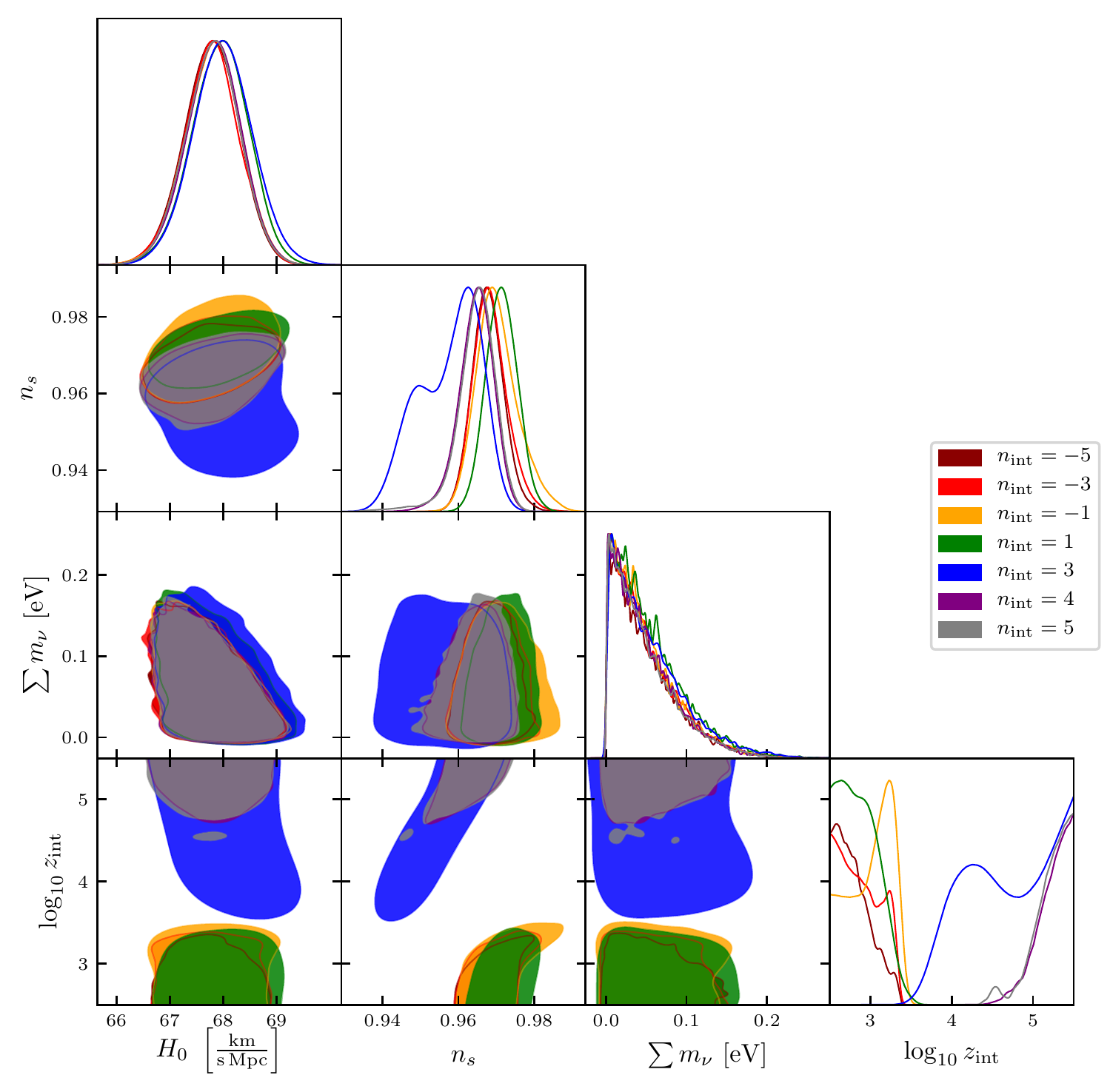}
\caption{Posterior probabilities from our analysis of Planck+BAO data for the parameters $H_0$, $n_s$, $\sum m_\nu$ and $\log_{10} z_{\rm int}$ for the 7 scenarios characterizing the suppression of neutrino freestreaming via a power-law $\Gamma_\text{nfs} \propto T^{n_{\rm int}}$ and with normalization such that $\Gamma_\text{nfs}(z_\text{int})=H(z_\text{int})$. In the 2-d contours we show $95\%$ C.L.\ regions. From the lower-right panel we can clearly appreciate a window of redshifts $2000\lesssim z_{\rm int} \lesssim 10^5$ where neutrino interactions are severely constrained. The only exception to this is the case $n_{\rm int }= 3$ which corresponds to a ratio $\Gamma_{\rm nfs}/H$ that is almost constant. This leads to modifications of the angular power spectra that are considerably degenerate with other cosmological parameters, in particular $n_s$, as shown in blue in this figure.}
\label{fig:posterior_nint_all}
\end{figure*}

\section{Planck constraints on neutrino freestreaming}\label{sec:results_models}

In this section we present Planck legacy constraints on the various rates described in Section~\ref{sec:methodology} that suppress neutrino freestreaming in the early Universe. For this purpose, we first implement the equations describing neutrino perturbations in the Boltzmann code {\tt CLASS}~\cite{Blas:2011rf,Lesgourgues:2011re}. Then, we perform a Markov-Chain-Monte-Carlo (MCMC) analysis using {\tt MontePython}~\cite{Audren:2012wb,Brinckmann:2018cvx}, for which we use Planck TT+TE+EE+lowE data~\cite{Planck:2018vyg} combined with data from Baryon Acoustic Oscillations (BAO) as in the  Planck legacy analysis~\cite{Beutler:2011hx,Ross:2014qpa,Alam:2016hwk}. We use the same priors as the Planck collaboration~\cite{Planck:2018vyg} for both the standard cosmological parameters as well as for the nuisance parameters in the Planck likelihoods. We use the following priors for the redshift at which the neutrino interaction goes above Hubble (see Eq.~\eqref{eq:Gamma_powerlaw}), which control the neutrino interaction rate in all the power-law cases:
\begin{align}
    \log_{10}z_{\rm int} &= [2,4]\,,\,\text{for}\,\,n_{\rm int } = [-5,\,-3,\,-1,\,1]\,,\\
    \log_{10}z_{\rm int} &= [3,6]\,,\,\text{for}\,\,n_{\rm int } = [3,\,4,\,5]\,.
    \label{eq:nint_prior}
\end{align}
For the transient scenarios, where the rate is parametrized by two parameters, $z_{\rm int}^{\rm max}$ and $\Gamma/H|_{\rm nfs}^{\rm max}$, we use the following priors:
\begin{align}
    \log_{10}z_{\rm int}^{\rm max} &= [1,7]\,, \\
    \log_{10}\Gamma/H|_{\rm nfs}^{\rm max} &= [-4,7] \,.
\end{align}
Note that the meaning of $z_{\rm int}$ and $z_{\rm int}^{\rm max}$ is different. While $z_{\rm int}$ refers to the time at which $\Gamma_{\rm nfs} = H$,  $z_{\rm int}^{\rm max}$ corresponds to the redshift at which the interaction rate divided by the expansion rate in the transient scenarios is largest. For instance, $z_{\rm int}^{\rm max} = 10^4$ for the interaction rate shown in orange in Fig.~\ref{fig:summary} while $z_{\rm int} = 10^5$ for the interaction rate in red in the same figure. For each interaction case our analysis chains consist of more than $3\times 10^{6}$ steps and we check that the Gelman-Rubin convergence diagnostic always satisfies $R - 1 < 0.02$ for all cosmological and nuisance parameters in the MCMC.

\begin{table}[t]
    \centering
{\def\arraystretch{1.55}
    \begin{tabular}{l|l|c|c}
    \hline\hline
     $\,\,\,$    Model $\,\,\,$ &  $\,\,\,$ Prior on $z_{\rm int}$    $\,\,\,$ & $\,\,\,$  $95\%$ C.L $\,\,\,$ & $\,\,\,$ $99.7\%$ C.L.  $\,\,\,$  \\ \hline \hline
$n_{\rm int} = -5 $ & $\log_{10} z_{\rm int} = [2,4]$ & $z_{\rm int} < 1300  $ & $z_{\rm int} < 2200   $ \\ \hline
$n_{\rm int} = -3 $ & $\log_{10} z_{\rm int} = [2,4]$ & $z_{\rm int} < 1900  $ & $z_{\rm int} < 2300   $ \\ \hline
$n_{\rm int} = -1 $ & $\log_{10} z_{\rm int} = [2,4]$ & $z_{\rm int} < 2400  $ & $z_{\rm int} < 3000   $ \\ \hline
$n_{\rm int} = 1 $  & $\log_{10} z_{\rm int} = [2,4]$ & $z_{\rm int} < 1800  $ & $z_{\rm int} < 3100   $ \\ \hline \hline
$n_{\rm int} = 3 $  & $\log_{10} z_{\rm int} = [3,6]$ & $z_{\rm int} > 6500 $  & $z_{\rm int} > 3300  $ \\ \hline
$n_{\rm int} = 4 $  & $\log_{10} z_{\rm int} = [3,6]$ & $z_{\rm int} > 92000 $ & $z_{\rm int} > 39000  $ \\ \hline
$n_{\rm int} = 5 $  & $\log_{10} z_{\rm int} = [3,6]$ & $z_{\rm int} > 85000 $ & $z_{\rm int} > 28000  $ \\ \hline \hline
    \end{tabular}
}
    \caption{Planck+BAO constraints on the redshift at which the rate suppressing neutrino freestreaming crosses Hubble, $z_{\rm int}$. The rates are described by $\Gamma_{\rm nfs} \propto T^{\rm n_{\rm int}}$. We also display the priors used for such parameter as well as the $95$\% and $99.7$\% C.L. bounds. The four upper rows correspond to interactions that become active at late times while the three lower rows correspond to the case of interactions being relevant at high redshift.}
    \label{tab:cosmo}
\end{table}

\subsection{The neutrino freestreaming window}

In Fig.~\ref{fig:posterior_nint_all} we display the posterior probabilities for $H_0$, $n_s$, $\sum m_\nu$ and $\log_{10} z_{\rm int}$ for the 7 power-law cases under study (a full $8\times 8$ plot is shown in Fig.~\ref{fig:posterior_nint_full}). Firstly, we do not find any relevant correlations between the interaction redshift and any standard cosmological parameter apart from the spectral index of primordial fluctuations, $n_s$, with the exception of the $n_{\rm int} = 3$ case (which is discussed separately below). In the scenarios with neutrinos interacting at high redshift the reconstructed value of $n_s$ is slightly smaller than in $\Lambda$CDM while for the case of neutrinos interacting at low redshift $n_s$ is slightly larger. Nevertheless, for any $n_{\rm int}\not=3$ the difference with respect to the $\Lambda$CDM value is within $\sim 1\sigma$ and therefore not too significant. This can be clearly be seen in Fig.~\ref{fig:posterior_nint_all} and we note that indeed similar trends were found in dedicated studies which focus solely on one type of interactions, see e.g.~\cite{Cyr-Racine:2013jua,Oldengott:2017fhy,Chen:2022idm,Forastieri:2019cuf}.  

Secondly, we do not find any statistically significant preference for neutrino interactions in any of these cases. As a result, we can obtain exclusion limits on $z_{\rm int}$ for the various power-law cases. In Tab.~\ref{tab:cosmo} we display the $95$\% C.L. and $99.7$\% C.L. limits on $z_{\rm int}$ for each scenario. We can clearly split the cases into two, the ones in which neutrinos interact at low redshift and in which they do so at high redshifts. In the low-redshift cases we find that the bound on $z_{\rm int}$ is somewhere between $1300-2400$. This result highlights the presence of a clear redshift window above which neutrinos cannot be interacting if the interaction kicks in at low-redshifts,
\begin{align}
    z_{\rm int} \lesssim 2000\,.
\end{align}
Similar constraints have been obtained in specific scenarios that consider one or another power-law index, see Refs.~\cite{Archidiacono:2013dua,Forastieri:2019cuf,Escudero:2019gfk,Chen:2022idm}. Importantly, because we consider all of them simultaneously we can see the moderate differences for each case. For example, comparing the $n_{\rm int}  = -5$ and $n_{\rm int} = -3$ cases we see that the bound on $z_{\rm int} $ is more stringent for the former. This was to be expected as the temperature dependence is sharper and does highlight that there lies some intrinsic uncertainty as to what the bound on $z_{\rm int}$ is, depending upon the power-law index that controls the rate.  Notice that the bound on $z_{\rm int}$ for the $n_{\rm int } = -1$ case is weaker than the rest because for this particular case we find a small $\sim 1\,{\sigma}$ preference for interactions. Preferences of similar strength in scenarios with neutrinos interacting at low temperatures have been reported in the literature~\cite{Archidiacono:2013dua,Forastieri:2019cuf,Escudero:2019gfk,Escudero:2019gvw,EscuderoAbenza:2020egd,Chen:2022idm,Venzor:2022hql}. As discussed in~\cite{Escudero:2019gfk} these weak preferences are likely driven by moderately low-$\ell$ Planck polarization data.

Next, looking at the scenarios in which neutrinos interact at high redshifts, we find 95\% C.L. bounds on $z_{\rm int}$ at the level of $z_{\rm int}> (8-9)\times 10^4$ for the cases $n_{\rm int} = 4,\,5$. The bound for $n_{\rm int } = 5 $,  $z_{\rm int} > 8.5\times 10^4$, can be compared with previous analyses in the literature, which found $z_{\rm int} \gtrsim 10^5$~\cite{Cyr-Racine:2013jua,Oldengott:2017fhy,Brinckmann:2020bcn}. Hence, our bound agrees very well with these previous results and the small discrepancy can largely be attributed to slight differences in the modeling and the parametrization/priors used to explore the parameter space (see also Section~\ref{sec:applications}). Taking into account the results from the power-laws with $n_{\rm int} = 4,\,5$, we see that Planck data restricts the interaction redshift to be 
\begin{align}
    z_{\rm int} \gtrsim 10^5\,,
\end{align}
for neutrinos interacting at high redshifts. 

The case $n_{\rm int} = 3$ is special as evident from Fig.~\ref{fig:posterior_nint_all}. The main reason for this is because the temperature dependence of the interaction rate  is not too different from that of the Hubble rate, $H\propto T^2$. This means that the CMB power spectrum in this particular scenario can be affected across a large range of multipoles. This in turn permits strong degeneracies in the fit allowing rather low $z_{\rm int}$ at the expense of a significant decrease of the spectral index $n_s$ as well as to some decrease of $A_s$ and a small increase of $H_0$. A degeneracy of this nature was already found in models with $n_{\rm int} =5$ before Planck legacy data arrived, see~\cite{Lancaster:2017ksf,Oldengott:2017fhy}. However, the case $n_{\rm int} =3$ allows for a much larger degeneracy because in these scenarios neutrino interactions alter a broader range of multipoles of which its effect can indeed be mimicked by shifts on $n_{s}$, $A_s$ and to some extent $H_0$. Importantly, since this affects $n_s$ and $A_s$ one should wonder whether Planck lensing data can break these degeneracies. For this purpose we have run an additional analysis including the lensing likelihood. However, we have actually found that the parameter space is unaltered by adding lensing to our baseline data sets. The interested reader can explicitly see this comparison in Appendix~\ref{sec:nint3appendix}.
 
Given these results what can we say about the neutrino freestreaming window? What we find is that for all scenarios in which neutrinos interact at low redshifts, $z_{\rm int} \lesssim 2000$, and for scenarios where the neutrinos interact at high redshift with rates that are significantly different than Hubble, $z_{\rm int } \gtrsim 10^5$. The case of $n_{\rm int} = 3$ represents a particular and curious case where degeneracies with $n_s$ and $A_s$ allow significantly smaller values of $z_{\rm int}$. In addition, we have found that Planck lensing data is not helpful in breaking this degeneracy and does not restrict further $z_{\rm int}$. 

We therefore conclude that there is a well defined neutrino freestreaming window $2000\lesssim z \lesssim 10^5$ where neutrinos should freestream in order to be compatible with Planck CMB observations. There is a small exception to this which corresponds to scenarios where neutrinos interact at high redshifts but where the rate suppressing neutrino freestreaming has a redshift dependence that is not too different from that of the Hubble rate -- here highlighted by the case $n_{\rm int} =3 $.

\subsection{Depth of the neutrino freestreaming window}

\begin{figure}[t]
\centering
	\includegraphics[width=0.46\textwidth]{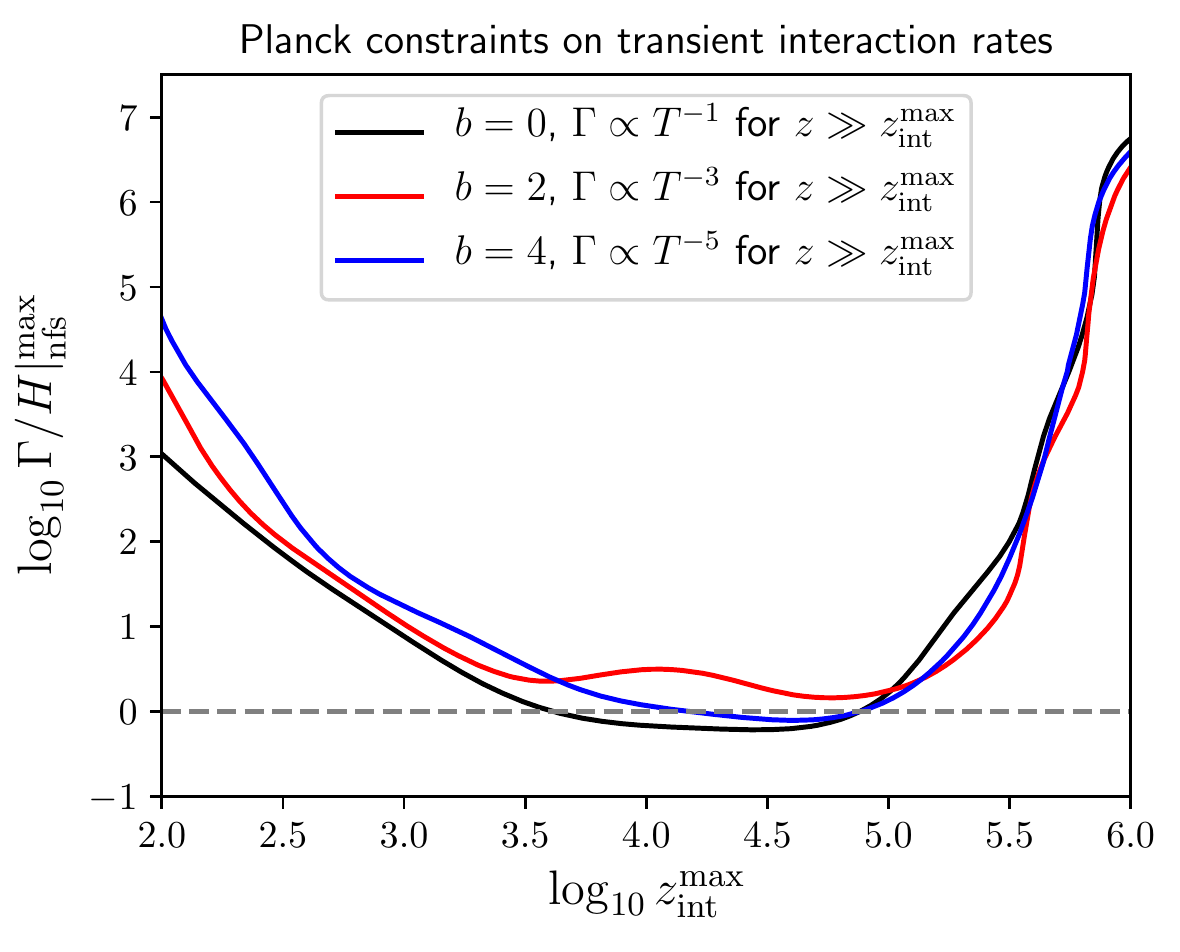}
\caption{Posteriors ($95\%$ C.L.) using Planck+BAO data for scenarios where the rate of neutrino freestreaming is transient, see orange line in Fig.~\ref{fig:summary}. In the y-axis we show the maximum value of $\Gamma_{\rm nfs}/H$ as a function of the redshift at which the rate becomes maxima, $z_{\rm int}^{\rm max}$. We show the results for the three cases $b = 0$, 2, 4, that differ in the high temperature limit $z\gg z_{\rm int}^{\rm max}$, namely corresponding to $\Gamma_\text{nfs} \propto T^{-1},\,T^{-3},\,T^{-5}$ respectively.}
\label{fig:b024_contours}
\end{figure}

The results from the power-law cases highlight the presence of a redshift window where neutrino freestreaming cannot be substantially suppressed, $ 2000 \lesssim  z_{\rm int} \lesssim 10^5$. However, on their own they cannot tell us the depth of this window, namely, they cannot tell us how large the interaction rate suppressing neutrino freestreaming can be within this window. In order to understand this we have run analyses for rates that are transient with three different high temperature dependencies as parametrized by the $b$ parameter in Eq.~\eqref{eq:bessel_int_improved}. The results for the Planck CMB analysis on the parameters characterizing these rates are shown in Fig.~\ref{fig:b024_contours}. We can appreciate that the results are fairly similar and a clear difference in the trend of these bounds appears only at $z_{\rm int}^{\rm max} \lesssim 10^3$. This was to some degree expected as for $z_{\rm int}^{\rm max} \lesssim 10^3$ the behavior of the tails of the rate are relevant, as controlled by the $b$ parameter.

\begin{figure}[t]
    \centering
	\includegraphics[width=0.48\textwidth]{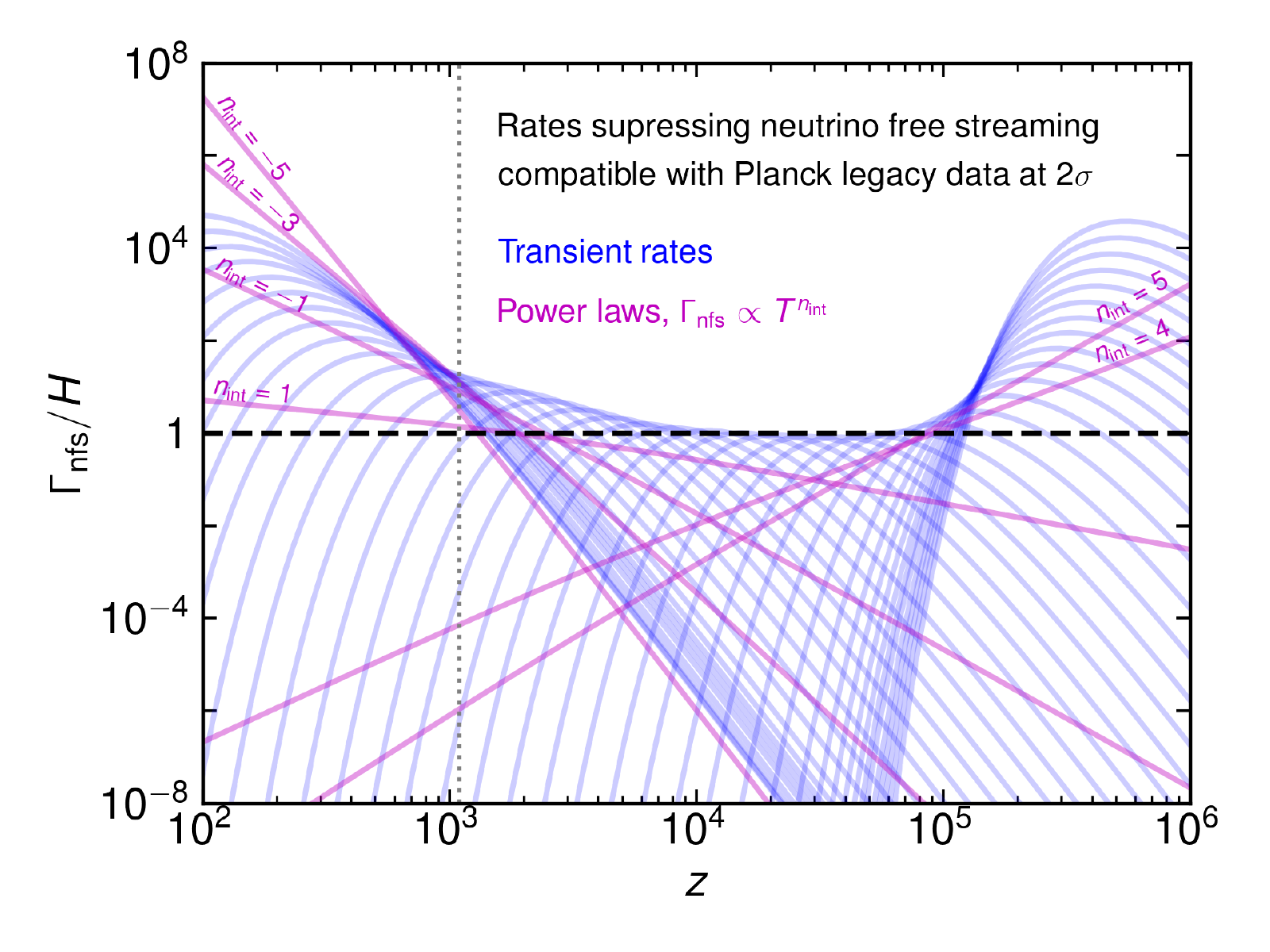}
    \caption{Rates suppressing neutrino freestreaming which are compatible at $2\sigma$ with Planck+BAO data. In magenta we show the results for the power-law cases, $\Gamma_{\rm nfs} \propto T^{n_{\rm int}}$. In blue we show transient rates. We have obtained these curves by taking points from the $2\sigma$ exclusion limit in Fig.~\ref{fig:b024_contours} for the $b= 4$ case. From this plot we can clearly appreciate the existence of a window in which neutrino freestreaming cannot be significantly modified, $10^3 \lesssim z_{\rm int} \lesssim 10^5 $, and also the strength of the interactions suppressing neutrino freestreaming within this window $\Gamma_{\rm nfs}/H\lesssim 1-10$.}
\label{fig:envelopes}
\end{figure}

An important feature of the bounds in Fig.~\ref{fig:b024_contours} is the flat region that covers roughly  $10^{3.5}\lesssim z_{\rm int}^{\rm max} \lesssim 10^5$. In addition, in this range the constraint for Planck+BAO on the parameter characterizing the maximum interaction strength is  $\Gamma/H|_{\rm nfs}^{\rm max} \lesssim 1-10$. This highlights that indeed, in this redshift window the interaction rate damping neutrino freestreaming  cannot  be too large even if the interaction is transient and is only effective for a limited period of time. 

\newpage 

From Fig.~\ref{fig:b024_contours} alone it is not possible to fully asses the shape of the window of redshifts in which neutrino freestreaming cannot be altered. In order to clearly see this we show in Fig.~\ref{fig:envelopes} the redshift evolution of rates suppressing neutrino freestreaming that saturate at $2\sigma$ the Planck+BAO bounds. In blue we show the results for transient scenarios. From these, we can appreciate certain interesting features. Firstly, we can see a high density of blue lines meeting in the region $z \sim (1-2)\times 10^5$. This clearly establishes the presence of this window and shows that neutrino freestreaming cannot be damped below such redshifts. This redshift roughly matches the one at which perturbations enter the horizon as relevant for CMB observations. Namely, $z\simeq 10^5$ corresponds to a conformal time $\tau \sim \tau_0/\ell$ with $\ell \sim 2000$ as relevant for Planck CMB observations.

In Fig.~\ref{fig:envelopes} we also show in magenta the redshift evolution of the power-law rates that saturate the $2\sigma$ Planck+BAO bounds. It is very interesting to see that the lines corresponding to interactions at high redshift meet at $z\sim 8\times 10^4$ and that they also meet with the envelope formed by the blue lines around that area.%
\footnote{We do not show the $n_{\rm int}= 3$ case because of the strong degeneracies that arise with $n_s$ and $A_s$ in this scenario.}
The region with $z \lesssim 10^4$ shows more features. It appears that the transient scenario does allow interactions to be efficient ($\Gamma/H\sim 1-10$) over some period of time close and prior to recombination. The reason for this may be related to the small preferences seen in scenarios with interactions of neutrinos at  low redshifts~\cite{Archidiacono:2013dua,Forastieri:2019cuf,Escudero:2019gfk,Escudero:2019gvw,Chen:2022idm,Venzor:2022hql}. Lastly, we see that at $z \sim 10^3$ the various power-laws and the transient rates globally meet. 

In summary, our Planck+BAO analysis highlights that neutrino freestreaming cannot be significantly suppressed in the redshift window $10^3 \lesssim z_{\rm int} \lesssim 10^5$, as can be most clearly seen in Fig.~\ref{fig:envelopes}. We have explicitly found that the exact boundary is largely model independent with the exception of the $n_{\rm int} = 3 $ case, where we find that degeneracies with $n_s$ and $A_s$ can allow neutrinos to be interacting at smaller redshifts than $10^5$. We summarize the precise bounds in Tab.~\ref{tab:cosmo} for the various power-law cases. With this window of redshifts in mind, we have subsequently performed analyses of rates where the interaction is transient, namely where one starts and ends up with freestreaming neutrinos but with a period in between where the anisotropic stress is damped. Our results for those types of scenarios show that the interaction rate as compared to Hubble cannot be significantly larger than $\Gamma_{\rm nfs}/H$ in this redshift window. Thus, we have delimited not only the extension of this window but also its depth.

\begin{figure*}[t]
\centering
	\includegraphics[width=0.32\textwidth]{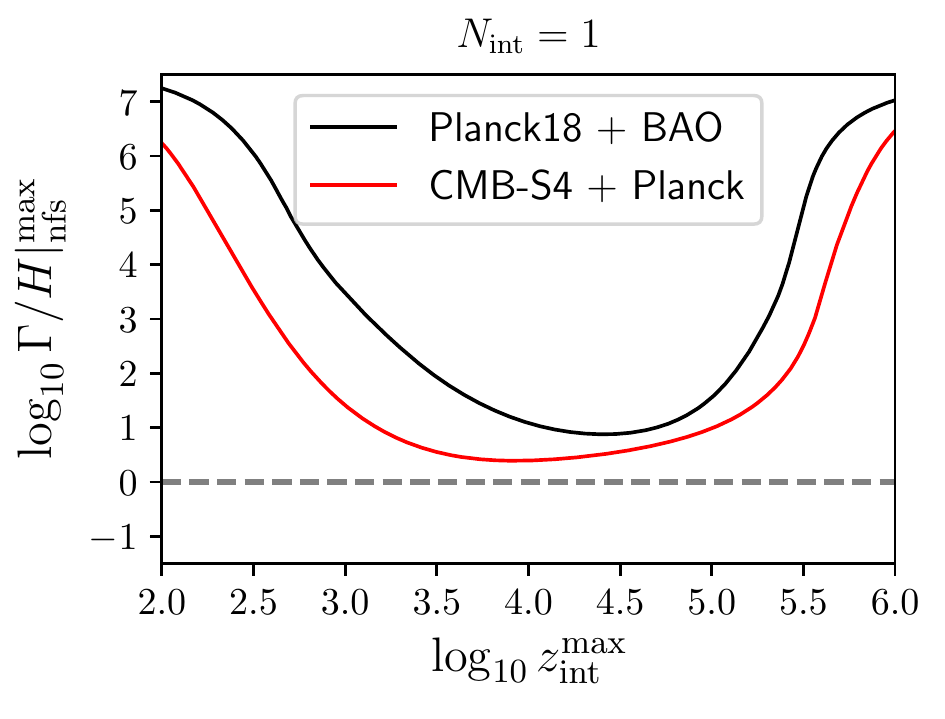}
	\includegraphics[width=0.32\textwidth]{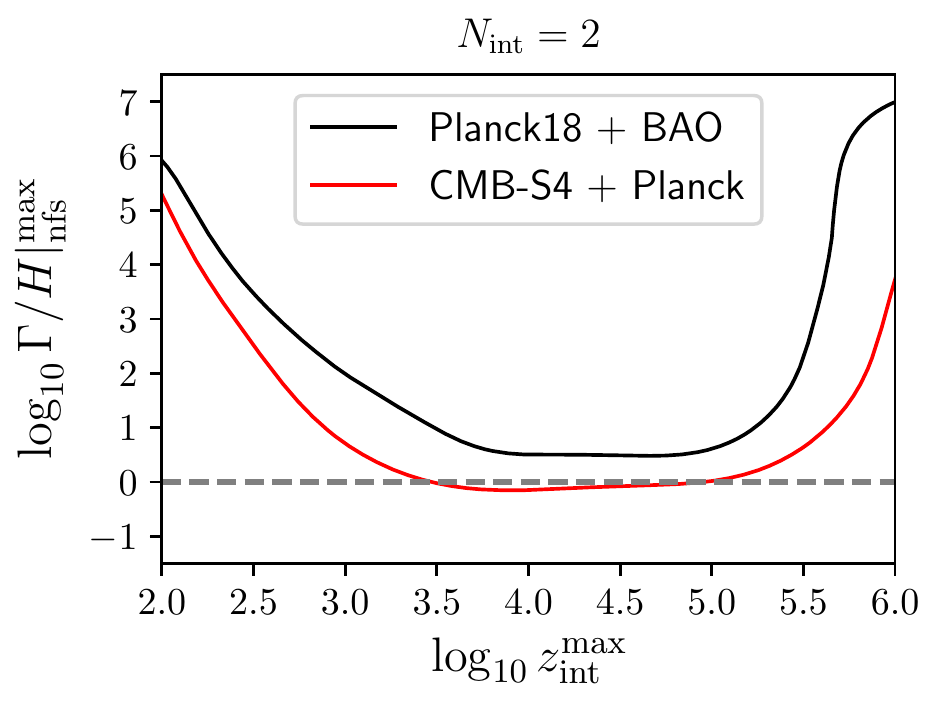}
	\includegraphics[width=0.32\textwidth]{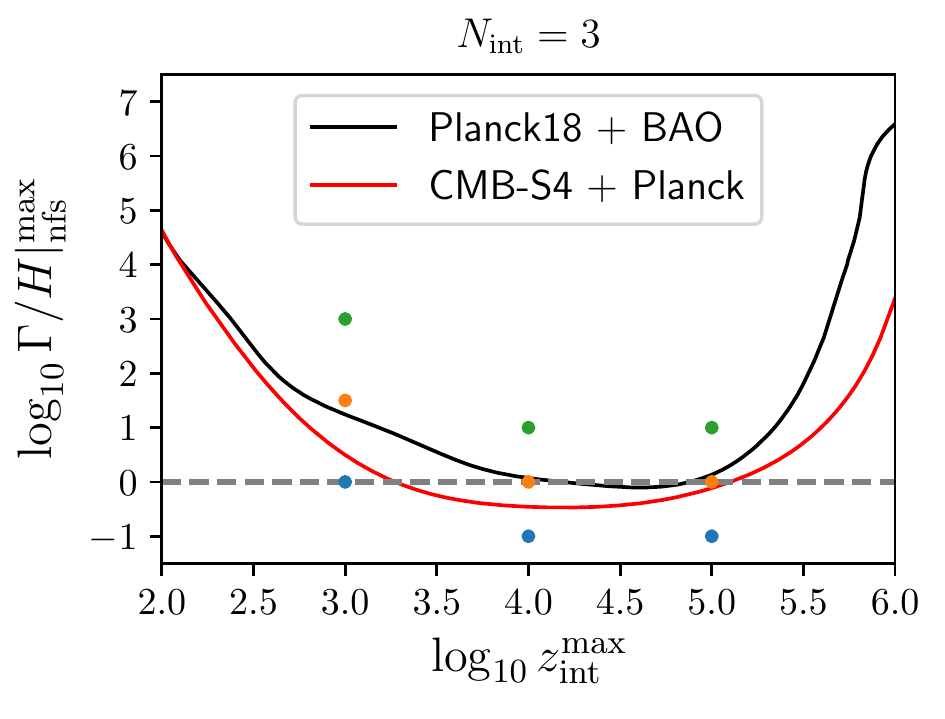}
\caption{$95\%$ C.L.\ upper limits for scenarios with a transient rate suppressing neutrino freestreaming. Each panel corresponds to the case of $N_{\mathrm{int}} = 1$, 2, and 3 interacting neutrinos, respectively. In black we show the current Planck+BAO bound and in red we show the forecasted reach for CMB-S4 in combination with Planck. The colored points in the rightmost panel correspond to a choice of parameters for which we explicitly show the angular power spectrum in Fig.~\ref{fig:cl_frac_diff}.}
\label{fig:CMBS4forecast}
\end{figure*}

\begin{figure*}[t]
\centering
	\includegraphics[width=0.32\textwidth]{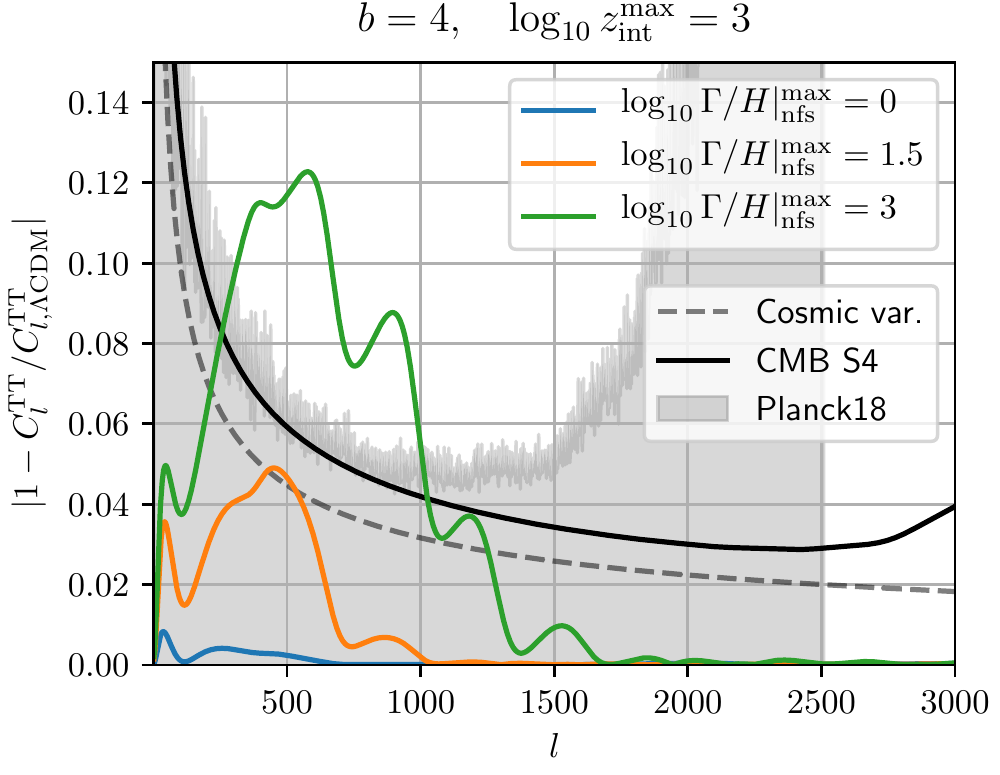}
	\includegraphics[width=0.32\textwidth]{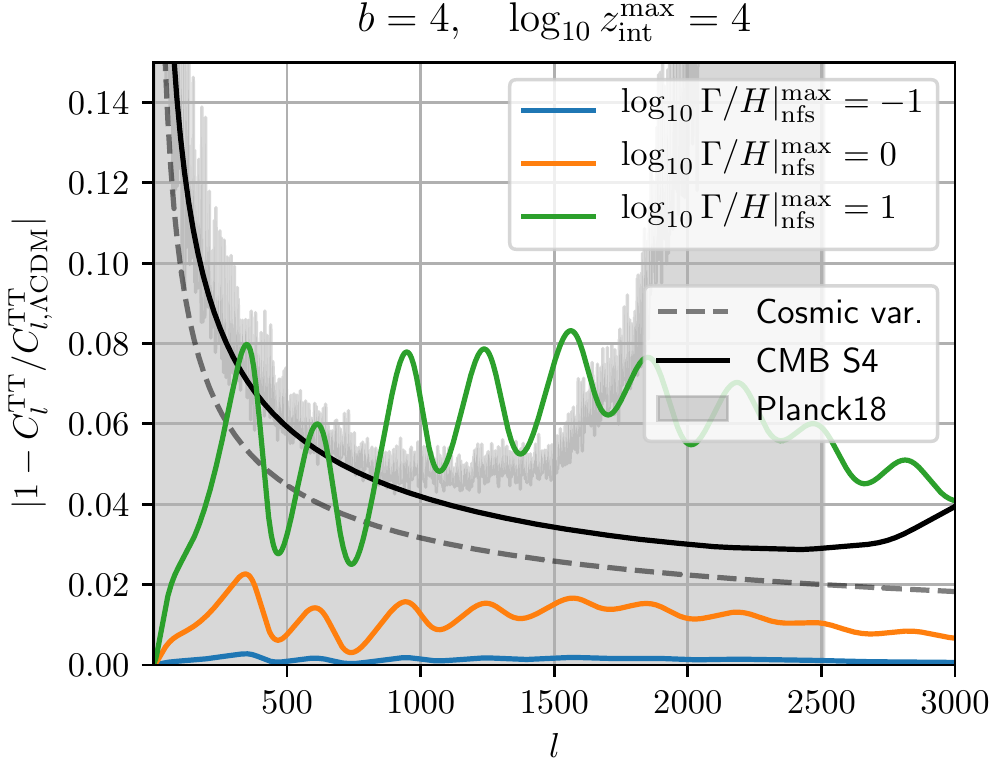}
	\includegraphics[width=0.32\textwidth]{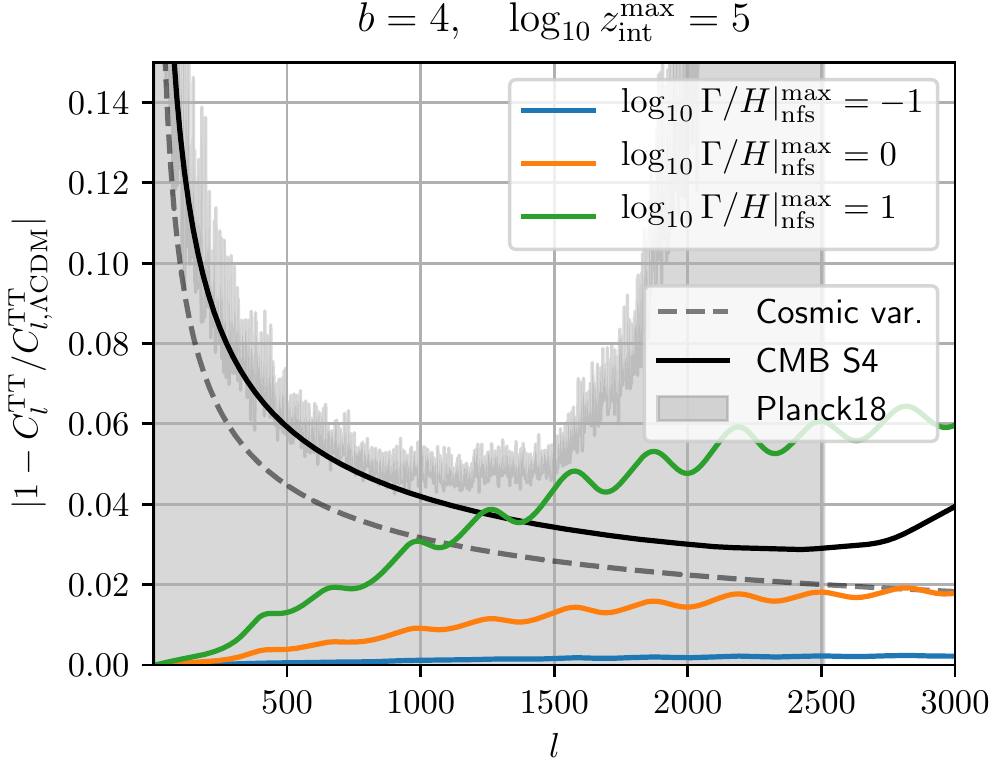}
\caption{Relative difference in the TT power spectra in the case of interacting neutrinos via a transient rate as compared to the \LCDM\ model (the fiducial cosmological parameters are fixed and we set $\sum m_{\nu} = 0.06~\eV$). We show the error bars from Planck CMB observations in grey, in black the expected sensitivity of CMB-S4 with $f_{\rm sky}$ = 0.57, and in dashed the cosmic variance limit. The parameters we pick for the interacting cases are given in the legend as well as displayed using the same colors in the rightmost panel of Fig.~\ref{fig:CMBS4forecast}.}
\label{fig:cl_frac_diff}
\end{figure*}

\newpage
\section{CMB Stage-IV results}\label{sec:CMBS4}
The next generation of CMB experiments, including the Simons Observatory~\cite{SimonsObservatory:2018koc}, LiteBIRD~\cite{LiteBIRD:2020khw}, and CMB-S4~\cite{Abazajian:2019eic}, are expected to significantly improve upon Planck's legacy by providing either better polarization measurements on large scales and/or by measuring in detail the small scale anisotropies of the CMB. In the context of neutrino freestreaming, we would like to understand the degree to which the next generation of CMB experiments can improve upon Planck on the region of redshifts at which neutrino freestreaming affects the CMB anisotropies. With this in mind, we run forecast analyses of CMB-S4 using again {\tt CLASS} and  {\tt MontePython}. To do this, we set up a futuristic likelihood assuming that the measured power spectrum would be given by a Planck-legacy cosmology and where the CMB likelihood is constructed using expected CMB-S4 error bars as done in~\cite{Brinckmann:2018owf}\footnote{In particular, we consider a data set combination of CMB-S4 for $f_{\rm sky} = 0.4$ for $50<\ell < 3000$ and a fake Planck Gaussian likelihood with $f_{\mathrm{sky}} = 0.57$ for $2< \ell <50$ and $f_{\mathrm{sky}} = 0.17$ for $51 < \ell < 3000$, for both temperature and polarization spectra. We used the inverse Fisher matrix as input covariance matrix to the MCMC runs, which yields a good guess for the covariance between the parameters and which significantly speeds up the sampling convergence~\cite{Brinckmann:2018cvx}.}.
In addition, in the previous section we have only considered cases where all three neutrino species were interacting. In order to see how much this assumption affects the window of redshifts where neutrino freestreaming cannot be altered, we consider in this section scenarios with $N_{\rm int} = 1,\,2,\,3$ interacting neutrinos. At the level of the MCMC analysis, we use the very same priors as in Section~\ref{sec:results_models}. 

In Fig.~\ref{fig:CMBS4forecast} we show the results for the case of neutrinos interacting with a rate that is transient as given in Eq.~\eqref{eq:bessel_int_improved} with $b = 4$. The three panels correspond to three cases where the number of interacting neutrino species is $N_{\rm int } = 1,\,2,\,3$, respectively. By looking at the Planck+BAO $2\sigma$ constraints we can appreciate that at a given $z_{\rm int}^{\rm max}$ the bounds on $\Gamma/H|_{\rm nfs}^{\rm max}$ become more stringent as the number of interacting neutrinos increases. This was clearly expected as the phase shift on the photon fluid generated by freestreaming neutrinos is directly related to the energy density they carry~\cite{Bashinsky:2003tk,Chacko:2003dt}.  In addition, we notice that the improvement of CMB-S4 over Planck is more significant for scenarios with $N_{\rm int} < 3$. 

By comparing in Fig.~\ref{fig:CMBS4forecast} the Planck+BAO and Planck+CMB-S4 we can clearly see that CMB-S4 observations have the power to test neutrino interaction rates that are $\sim 1$ order of magnitude weaker than Planck, in general. The expected improvement from CMB-S4 depends sensitively on the value of $z_{\rm int}^{\rm max}$ and the number of interacting neutrino species $N_{\rm int}$.

In general, when increasing $z_{\rm int}^{\rm max}$ the effect of neutrino interactions on the angular power spectrum shifts towards higher $\ell$. This is because interactions were active at earlier times, corresponding to smaller scales. Due to the increase in sensitivity of CMB-S4 on small angular scales, the improvement of the bound on the interaction strength is particularly strong for large $z_{\rm int}^{\rm max}$. Note that this argument applies directly to $N_{\rm int}=3$. For $N_{\rm int}=1$, and to a certain extent also $N_{\rm int}=2$, a CMB-S4 type experiment can improve the sensitivity to neutrino interactions considerably also for low values of $z_{\rm int}^{\rm max}$. This is due to the larger allowed values of
the maximal interaction strength $\Gamma/H|_{\rm nfs}^{\rm max}$ when only a single neutrino species is interacting as compared to the case where all of them are. Hence, the interaction can be active over a longer period of time for $N_{\rm int}=1$ as compared to $N_{\rm int}=3$. For a given $z_{\rm int}^{\rm max}$ the impact on the angular power spectrum is therefore also shifted to higher $\ell$ when lowering $N_{\rm int}$. This explains the increase in sensitivity of CMB-S4 over Planck for low $z_{\rm int}^{\rm max}$ and $N_{\rm int}=1$.

In order to have an understanding of the bounds and expected sensitivity shown in Fig.~\ref{fig:CMBS4forecast}, we show in Fig.~\ref{fig:cl_frac_diff} the relative difference of the TT power spectra for three values of $z_{\rm int}^{\rm max}$ for representative values of $\Gamma/H|_{\rm nfs}^{\rm max}$ as highlighted with dots in the right panel of Fig.~\ref{fig:CMBS4forecast}. In these plot we show the size of Planck error bars as well as those expected for CMB-S4~\cite{Brinckmann:2018owf}. We have chosen these points to highlight regions of parameter space which are clearly excluded by Planck data (green), at the boundary of Planck constraints but within the reach of CMB-S4 data (orange), and those that cannot be probed even with a cosmic variance limited experiment (in blue). 

In addition, we have run CMB-S4 forecasts for a power-law rate with power $n_{\rm int} = 3$ and $n_{\rm int} = 5$, as we expect CMB-S4 to be most sensitive to cases where neutrinos interact at high redshift. In particular, we find that CMB-S4 in combination with Planck will have a sensitivity to neutrino interactions up to redshift
\begin{align}
  \!\! n_{\rm int} &= 5\,\,\,\,\Longrightarrow  \,\,\, z_{\rm int} < 2.8\times 10^5\,(2\sigma)\,\,\,[\text{CMB-S4}]\,,\\
   \!\!  n_{\rm int} &= 3\,\,\,\,\Longrightarrow  \,\,\, z_{\rm int} < 2.4\times 10^5\,(2\sigma)\,\,\,[\text{CMB-S4}]\,.
\end{align}
For the case of $n_{\rm int} =5$ this represents an improvement of a factor $\sim 3$ to the current level of sensitivity that Planck observations have, see Tab.~\ref{tab:cosmo}. In this case, the improvement in sensitivity arises from the fact that CMB-S4 should be able to measure the CMB anisotropies with almost cosmic variance error up to $\ell \simeq 3000$. Since perturbation modes with higher $\ell$ enter the horizon earlier, CMB-S4 is sensitive to neutrino interactions at higher redshift as compared to Planck.

The case of $n_{\rm int} =3$ is particularly interesting in light of the results we found in Section~\ref{sec:results_models}, namely, that rather small values of $z_{\rm int}$ are allowed thanks to a degeneracy of this rate with $n_s$ and to some degree with $A_s$ and $H_0$. The upcoming sensitivity of CMB-S4 is quite telling, it shows that CMB-S4 will be able to break this degeneracy  (likely from cosmic variance measurements of the EE spectrum). This would be important because it could shape the neutrino freestreaming window  even for the case $n_{\rm int} = 3$ up to $z_{\rm int} \sim 2\times 10^5$. 

In summary, we have explored the reach of CMB Stage-IV experiments to neutrino freestreaming in the early Universe. Our analysis shows that in the cases where the rate that damps neutrino freestreaming is transient, CMB-S4 would be sensitive to interaction rates that are roughly an order of magnitude smaller than those that are currently probed by Planck. In addition, CMB-S4 observations will be able to test neutrinos interacting up to a redshift $z_{\rm int} \lesssim 3 \times 10^5$. 

\section{Implication for LSS observations}\label{sec:LSS}

\begin{figure*}[t]
\centering
	\includegraphics[width=0.32\textwidth]{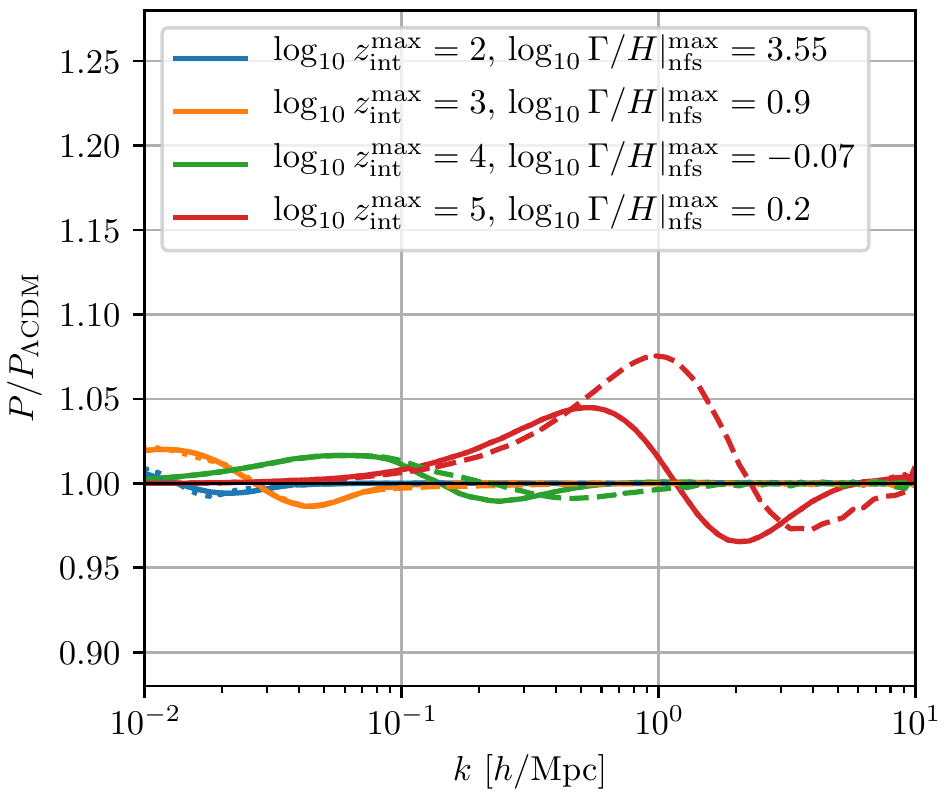}
	\includegraphics[width=0.32\textwidth]{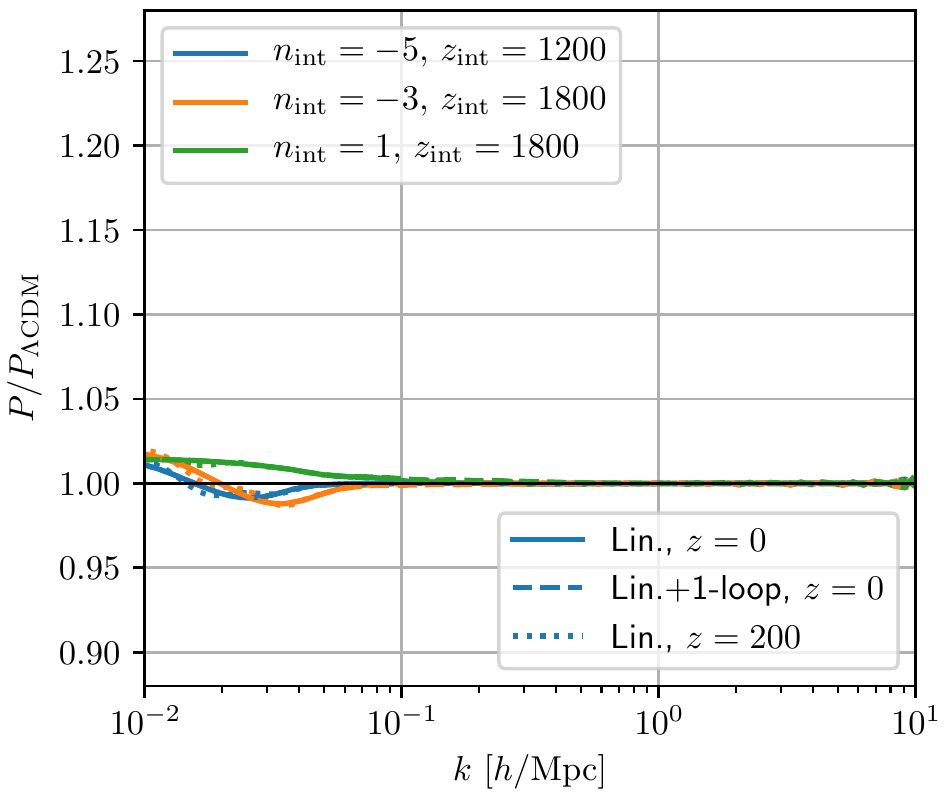}
	\includegraphics[width=0.32\textwidth]{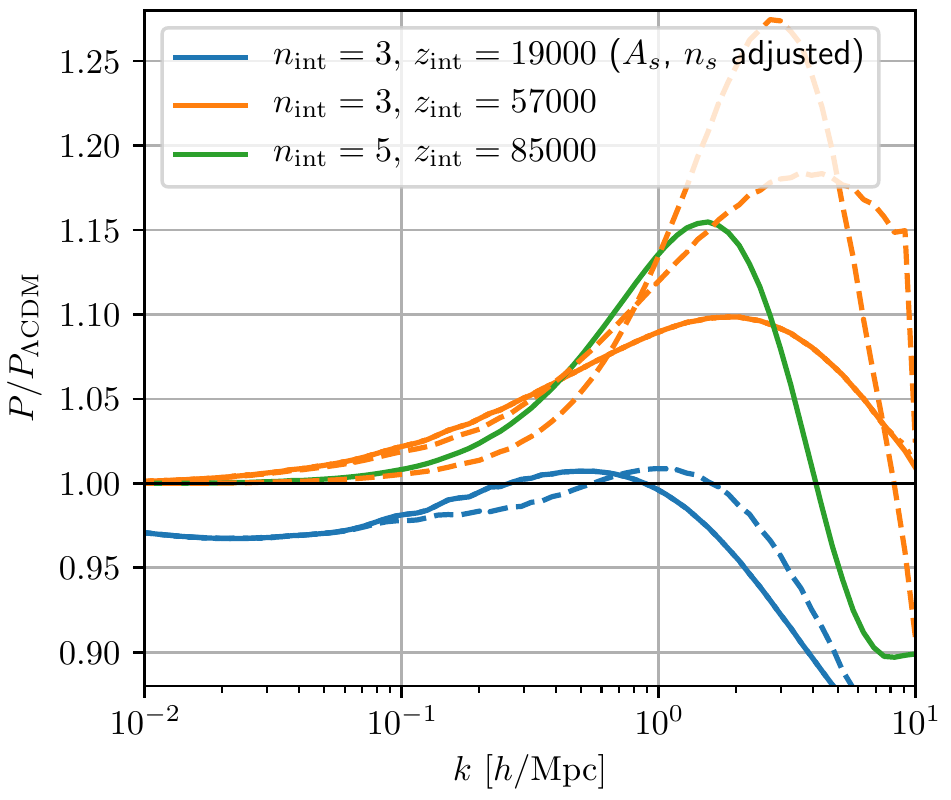}
\caption{The power spectrum in models with interacting neutrinos normalized to that in a $\Lambda$CDM model (with massive neutrinos $\sum m_{\nu} = 0.06~\eV$).
We show the linear (solid) and linear + one-loop (dashed) spectra at $z = 0$ and the linear spectrum at $z = 200$ (dotted, mostly invisible behind the solid curves).
\emph{Left:} Transient interaction case with $b=4$ and various amplitudes $\Gamma/H|_{\mathrm{nfs}}^{\mathrm{max}}$.
\emph{Middle:} Power-law cases interacting at low redshift, in particular $\nint = -5$ (blue), $\nint = -3$ (yellow) and $\nint=1$ (green).
\emph{Right:} Power-law cases $\nint = 5$ (blue) and $\nint = 3$ (yellow) interacting at high redshift. For $\nint = 3$ we display also a case with an intermediate interaction redshift $\zint = 19000$ (blue), with reduced $A_s = 1.995\times 10^{-9}$ and $n_s = 0.953$. The $\Lambda$CDM model that we compare to has $A_s = 2.116\times 10^{-9}$ and $n_s = 0.966$.
}
\label{fig:pk_frac_diff}
\end{figure*}

Large-scale structure surveys mapping the distribution of tracers of the matter density field are sensitive to modifications of the matter power spectrum on weakly non-linear scales $k\sim {\cal O}(0.1) h/\text{Mpc}$, within the regime of baryon acoustic oscillations. Ongoing and future galaxy surveys such as DES~\cite{DES:2021wwk} and DESI~\cite{Aghamousa:2016zmz} as well as the Vera Rubin Observatory~\cite{2019ApJ...873..111I} and the satellite telescopes Roman~\cite{2021MNRAS.tmp.1608E} and Euclid~\cite{Amendola:2016saw} will increase the sensitivity to the (few-)percent level, building on the eBOSS legacy~\cite{Gil-Marin:2018cgo,Bautista:2020ahg}.
Furthermore, DESI collects a large number of quasar absorption spectra that will allow us to probe the matter power spectrum via the Lyman-$\alpha$ forest~\cite{Karacayli:2020aad} on smaller scales $k\sim {\cal O}(1) h/\text{Mpc}$ (around the non-linear scale at the observed redshifts $z\sim 2-4$), significantly extending the latest BOSS DR14 data set~\cite{Chabanier:2018rga}.

Neutrino interactions can potentially influence the matter power spectrum on these scales mainly via two effects:
\begin{itemize}
    \item Massive neutrinos lead to a well-known suppression of the matter power spectrum, originating from a slower growth of matter perturbations on scales with wavenumber $k\gg k_\text{fs}$ relative to the case of massless neutrinos, where  $k_\text{fs}\sim 0.05h/\text{Mpc}(m_\nu/0.1\text{eV})(\Omega_m^0/0.3)^{1/2}/(1+z)^{1/2}$ is the freestreaming scale, while behaving as cold dark matter on large scales, i.e.\ for $k\ll k_\text{fs}$. Neutrino interactions at low redshift $z\ll z_\text{nr}$ alter the anisotropic stress, that in turn affects the evolution of neutrino perturbations, and can modify the scale-dependence of the suppression of the matter power spectrum  on scales $k\sim k_\text{fs}$.
    \item Neutrino interactions at high redshift, i.e.\ for $z\gtrsim {\cal O}(z_\text{eq}=3400)$, alter
    the gravitational potentials and therefore also the (logarithmic) growth of matter perturbations during the radiation era. This imprints a modification on the matter power spectrum for modes entering the horizon in these epochs, and while neutrino interactions are relevant, analogously to the CMB spectra.
\end{itemize}
In Fig.\,\ref{fig:pk_frac_diff} we show the relative difference $P(k)/P_{\Lambda\text{CDM}}(k)$ of the matter power spectrum for the case with and without interactions, for the cases of a transient interaction rate (left), power-law interaction active at low redshift (middle) and at high redshift (right).
In order to assess the maximal effect consistent with CMB constraints, we choose an interaction strength that is close to the boundary of the $95\%$ C.L.\ allowed region of our CMB analysis. In order to disentangle the two effects mentioned above, we show the ratio of linear power spectra today ($z=0$, solid lines) and at the time when neutrinos start to become non-relativistic ($z=200$, dotted lines). We find practically no difference, implying that the second effect discussed above is by far dominant. 

The dominance of the second mechanism implies that the imprint on the power spectrum is much larger for scenarios where neutrinos interact at high redshift, as can be seen by comparing the middle and right panels of Fig.\,\ref{fig:pk_frac_diff} for power-law cases, as well as by comparing the results for the various $z_\text{max}$ in the left panel. For example, for transient interactions with $z_\text{max}=10^5$ the linear power spectrum is modified at the $5\%$ level, and for power-law interactions with $n_\text{int}=5$ at the $10- 15\%$ level. The scales at which these deviations occur correspond to the perturbation modes that enter the horizon while the damping rate $\Gamma_\text{nfs}$ of anisotropic stress due to neutrino interactions is comparable to or larger than the Hubble scale. 

In order to illustrate the relevance of non-linear effects we also show the impact of adding the first non-linear correction within perturbation theory for large-scale structure~\cite{Bernardeau:2001qr}, the so-called 1-loop correction, by the dashed lines in Fig.\,\ref{fig:pk_frac_diff} (see~\cite{Fuhrer:2014zka,Garny:2020ilv,Chen:2020bdf, Garny:2022fsh} for schemes taking the neutrino anisotropic stress into account beyond the linear approximation). The 1-loop correction is evaluated at $z=0$, and we observe an enhancement of the differences compared to the non-interacting case. Nevertheless, on scales where the solid and dashed lines deviate significantly, a full non-linear analysis would be required. Note that the size of the loop correction roughly decreases with the second power of the linear growth function at higher redshift relative to the linear contribution, such that non-linear corrections are smaller by a factor $\sim\! 10$ at redshifts $z=2-4$.

The change of the matter power spectrum is most pronounced for modes of order $1-10h/\text{Mpc}$ in scenarios with interactions in the pre-recombination era. Therefore, future Lyman-$\alpha$ observations by DESI are a promising strategy to further test this scenario, and, depending on the assumptions on astrophysical uncertainties~\cite{Garzilli:2021qos}, even current BOSS data may already be competitive to CMB bounds~\cite{Chabanier:2018rga}. In particular, the pronounced $k$-dependence of the impact of neutrino interactions on the matter power spectrum provides a promising feature to break degeneracies with astrophysical parameters of the intergalactic medium~\cite{Garny:2020rom}. For galaxy surveys it will depend on the ability to disentangle galaxy bias from changes of the underlying matter power spectrum whether neutrino interactions can be tested~\cite{Boyle:2020rxq,Pezzotta:2021vfn}. 

Note that for the case of $n_{\rm int} = 3$, which is somewhat special as discussed above, we show power spectra for two benchmark scenarios in the right panel of Fig.\,\ref{fig:pk_frac_diff}. Both of them are compatible with current Planck constraints. The first one has a relatively strong interaction, that is still allowed by Planck due to a degeneracy with $n_s$ and $A_s$. Thus, for the first model, we adjusted $n_s$ and $A_s$ accordingly. Therefore, the power spectrum deviates from the Planck best-fit $\Lambda$CDM model even on very large scales, at the few percent level. In addition, it is suppressed on small scales. Consequently, the degeneracy may be broken by upcoming DESI Lyman-$\alpha$ and possibly galaxy clustering data, apart from CMB-S4 experiments in the future, as discussed above. The second benchmark model with $n_{\rm int} = 3$ has a weaker interaction strength (corresponding to larger $z_{\rm int}$), for which $n_s$ and the other cosmological parameters coincide with those of the Planck best-fit $\Lambda$CDM model. Nevertheless, it leads to deviations in the power spectrum at the $10\%$ level on scales that are going to be probed by DESI.

We conclude that neutrino interactions in the pre-recombination era can be tested by DESI via Lyman-$\alpha$ forest observations, while neutrino interactions in the post-recombination era are much harder to constrain with future large-scale structure observations. In particular, there are good prospects to break the degeneracy that occurs for Planck observations in the  $n_{\rm int} = 3$ case with large-scale structure data.

\section{Neutrino Interaction Models and comparison with the literature}\label{sec:applications}

The aim of this section is threefold. Firstly, we want to highlight how the phenomenological rates suppressing neutrino freestreaming used in the previous sections can be related to particle physics models of interacting neutrinos. Secondly, we use these relations to see how our direct bounds on the redshift at which the neutrinos stop to freestream, $z_{\rm int}$, can be mapped onto bounds on particle physics parameters such as couplings and masses of new states. Thirdly, we compare these bounds to dedicated studies in the literature that focus on specific scenarios. This will also serve as a test of the approximations used in our modelling as discussed in Section~\ref{sec:methodology}.

Globally, we aim to demonstrate how the model-independent approach pursued in this work can be applied to obtain bounds on parameters within specific models, taking as an example a number of well-known cases. This is intended to serve as an instruction on how to use our results for estimating bounds in scenarios beyond the Standard Model that are yet to be developed.

\textit{--- $\nu \nu \leftrightarrow \nu\nu$: Neutrino-neutrino annihilations and scatterings.} Neutrino self-interactions in cosmology have been a topic of intense study~\cite{Cyr-Racine:2013jua,Oldengott:2014qra,Lancaster:2017ksf,Oldengott:2017fhy,Kreisch:2019yzn,Park:2019ibn,Das:2020xke,RoyChoudhury:2020dmd,Brinckmann:2020bcn,Kreisch:2022zxp}. On dimensional grounds, the rate at which the neutrino anisotropic stress is damped is expected to be $\Gamma_{\rm nfs} \sim G_{\rm eff}^2 T^5$ where $G_{\rm eff}$ is an effective Fermi-constant parametrizing these interactions. In this context, the authors of Ref.~\cite{Oldengott:2014qra,Oldengott:2017fhy,Kreisch:2019yzn} have explicitly calculated the resulting inhomogeneous Boltzmann equations for this scenario. The actual result can be put in a similar form to our Eqs.~\eqref{eq:Boltzmannhierarchy}, but with some small $\ell$ dependence for the terms $\ell \geq 2 $ which we do not account for. However, this $\ell$ dependence has been shown to be less than 20\% between $\ell = 2$ and $\ell = 6$, see Eq.~(2.9) in Ref.~\cite{Oldengott:2017fhy}. This implies that one can in practice actually treat it as $\ell$ independent and therefore in this case $\Gamma_{\rm nfs} \simeq  G_{\rm eff}^2 T_\nu^5$. In our notation, it corresponds to the power-law case of $n_{\rm int} = 5$. Translating the bound from Tab.~\ref{tab:cosmo} of $z_{\rm int}>8.5\times 10^4 $ at $95$\% C.L.\ yields a constraint of $G_{\rm eff} < 4.1\times 10^{-4}\,{\rm MeV}^{-2} = 3.5\times 10^7\,G_F$. This number can be compared to the $95$\% upper limit from~\cite{RoyChoudhury:2020dmd} that was derived using the same data sets we consider here (see their Tab. 2): $G_{\rm eff} < 4.3\times 10^{-4}\,{\rm MeV}^{-2}$. Therefore, we see that the results are in excellent agreement. On another note, strong neutrino self-interactions of this type were proposed as a possible avenue to solve the Hubble tension~\cite{Kreisch:2019yzn}. However, it was later shown that the solution does not stand when Planck-legacy polarization data is used~\cite{Das:2020xke,RoyChoudhury:2020dmd,Brinckmann:2020bcn}. Here, we consider Planck legacy polarization data and we do not find any relevant preference for interactions either, as can be seen from the posterior for $z_{\rm int}$ in the case of $n_{\rm int} = 5$ in Fig.~\ref{fig:posterior_nint_all}. We note that, very recently, a study of this scenario using data from the Atacama Cosmology Telescope (ACT)~\cite{ACT:2020gnv} shows a 2-3$\sigma$ preference for neutrino interactions~\cite{Kreisch:2022zxp}. When Planck data is included in the analysis, however, this preference is reduced.  

\textit{--- $\nu \nu \leftrightarrow \phi\phi$: Neutrino annihilations into massless scalars.} Another scenario studied in the literature is the case of neutrinos annihilating into massless species~\cite{Beacom:2004yd,Hannestad:2004qu,Bell:2005dr,Archidiacono:2013dua,Forastieri:2015paa,Forastieri:2019cuf,Venzor:2022hql}. The exact Boltzmann equation has been formally derived in~\cite{Oldengott:2014qra}, but has so far never been solved explicitly in the literature. The most recent analyses of this scenario are performed in Refs.~\cite{Forastieri:2015paa,Forastieri:2019cuf,Venzor:2022hql} and use a relaxation time approximation for the collision term and assume a coupled $\nu\!-\!\phi$ system to model the perturbations. This effectively means that the damping of the neutrino anisotropic stress is considered to be proportional to the energy transfer rate which is precisely what we assumed in Eq.~\eqref{eq:psil}. In order to map to a region of parameter space we need to consider an explicit model. For illustration purposes, we consider Majorana neutrinos coupled to a light pseudoscalar $\phi$ with an interaction Lagrangian $\mathcal{L}_{\rm int} =\sum_{i=1}^3  \,(\lambda/2) \,\bar{\nu}_i\gamma_5\nu_i \phi$, where $\lambda$ is a dimensionless coupling constant. In this case, the annihilation cross section between two massless scalar particles and two neutrinos is $\sigma(s) \simeq \lambda^4/(32\pi s) \log ( s/m_\nu^2)$~\cite{EscuderoAbenza:2020cmq,Escudero:2019gfk} where $\sqrt{s}$ is the center of mass energy. This means that the rate suppressing neutrino freestreaming can be parametrized as $\Gamma_{\rm nfs} \simeq 10^{-3} \, \lambda^4 T_\nu $. We have obtained this by taking Eq.~(A4) of \cite{Escudero:2019gfk} and dividing it by $\rho_\phi = T_\nu^4 \pi^2/30$, neglecting a small logarithmic correction to this formula. This is precisely of the form of our power-law with index $n_{\rm int} = 1$ for which we obtained $z_{\rm int} < 1800$ (see Tab.~\ref{tab:cosmo}). This in turn implies a bound on the coupling $\lambda \lesssim 7.1\times 10^{-7}$ at 95\% C.L. While Ref.~\cite{Forastieri:2019cuf} uses a slightly different effective coupling we have explicitly checked that their resulting bound can be translated into $\lambda < 7.4\times 10^{-7} $ using Planck 2015+BAO data. We thus see a very similar result and we attribute the slight improvement of the bound to updated Planck polarization data in 2018 as compared to 2015.

\textit{--- $\phi \leftrightarrow \bar{\nu}\,\nu$: eV-scale neutrinophilic bosons.} Decays and inverse decays of eV-scale neutrinophilic bosons ($\phi$) were the first type of interacting neutrino scenario considered in the context of CMB constraints~\cite{Chacko:2003dt}. The Boltzmann hierarchy as relevant for neutrino freestreaming for this scenario has only been recently explicitly computed in Refs.~\cite{Barenboim:2020vrr,Chen:2022idm}. These references have shown that actually the naive relaxation time approximation is not a good ansatz for the rate that suppresses neutrino freestreaming. In particular, they have shown that the rate suppressing neutrino freestreaming in these scenarios should be $\Gamma_{\rm nfs} \simeq \Gamma_\phi\, (m_\phi/T)^5$ at $T\gg m_\phi$ and $\Gamma_{\rm nfs}\simeq \Gamma_\phi\, e^{-m_\phi/T}$ at $T\ll m_{\phi}$, where $\Gamma_\phi$ is the decay rate in vacuum of the $\phi$ particle into neutrinos. This is substantially different to what one would naively expect at $T\gg m_\phi$ from the typical energy transport rate for decays and inverse decays $\Gamma \simeq \Gamma_\phi\, (m_\phi/T)$~\cite{Kolb:1990vq}, or the one expected by taking a random walk at the background level, $\Gamma_{\rm nfs} \simeq \Gamma_\phi\, (m_\phi/T)^3 $~\cite{Chacko:2003dt,Hannestad:2005ex}. 

In this context, the only CMB analyses of this scenario presently in the literature used the naive transport rate for decays and inverse decays~\cite{Escudero:2019gvw,Escudero:2021rfi}. While clearly there is a difference in scaling, the difference between all these cases is actually not too significant when the rate is maximal, which corresponds to $T\sim m_\phi/3$. Around and below these temperatures all of these rates roughly match. In any case, taking the formulae in~\cite{Barenboim:2020vrr,Chen:2022idm} (see in particular Eq.~(13) of~\cite{Chen:2022idm}) one can do an approximate mapping into the decay rate in vacuum of a neutrinophilic boson with masses below the keV scale. For concreteness, one can consider the case of a scalar $\phi$ as the one considered in the previous paragraph again described by $\mathcal{L}_{\rm int} =\sum_{i=1}^3  \,(\lambda/2) \,\bar{\nu}_i\gamma_5\nu_i \phi$ but now with $m_\phi > 2m_\nu$. In this case, the process of decays and inverse decays will form a coupled $\nu\!-\!\phi$ system where the $\phi$ component represents $\sim 10\% $ of the energy density~\cite{EscuderoAbenza:2020cmq} (provided that there is no primordial population of these species). In this scenario, one can relate the rate in Eq.\ 13 of~\cite{Chen:2022idm} to our rate in Eq.~\eqref{eq:bessel_int_improved} by doing the following mapping: $z_{\rm int}^{\rm max} \simeq  1200 \,m_\phi/\,{\rm eV}$ and by relating $\Gamma/H|_{\rm nfs}^{\rm max} \simeq \Gamma_\phi/(80 H(z_{\rm int}))$ for the $b = 4$ scenario. For these settings our rate agrees with that in~\cite{Chen:2022idm} by better than a factor of $2$ for $0.1 < T_\gamma/m_\phi < 10$, which is the relevant range of temperatures (namely, this is when the rate can become large as compared to $H(z)$). With this in mind, having a look at the black contour in Fig.~\ref{fig:b024_contours} we can see that bosons with a mass $0.1\,{\rm eV} \lesssim m_\phi \lesssim 200\,{\rm eV}$ can in principle be constrained by CMB observations. For illustration purposes we can consider a given point. Taking the bound of $\Gamma/H|_{\rm nfs}^{\rm max} \gtrsim 1$ at  $z_{\rm int}^{\rm max} \simeq 10^4$ (see Fig.~\ref{fig:b024_contours}) then one can bound the lifetime of such a scalar of mass $m_\phi \simeq 10\,{\rm eV}$ to be $\Gamma_\phi < 80\,H(z=10^4)$. In this case, the lifetime can be written as $\Gamma_\phi = 3 m_\phi \lambda^2/(16\pi) $. Rewriting this bound in terms of the coupling we obtain $\lambda \lesssim 5\times 10^{-13}$ at $2\sigma$. One can compare this bound with the results in Refs.~\cite{Escudero:2019gvw,Escudero:2021rfi} which performed a Planck analysis but using the other energy transport rate which for such mass find $\lambda\lesssim 2\times 10^{-13}$. We clearly see that there is an overall agreement but a relaxation of the bound by a factor of $\sim 3$ as a result of the fact that the actual rate reducing neutrino freestreaming calculated in~\cite{Barenboim:2020vrr,Chen:2022idm} is smaller than the one considered in~\cite{Escudero:2019gvw,Escudero:2021rfi}. 

Finally, there is a small caveat for the bounds in these types of scenarios. The reason is that in this case one does expect a potentially relevant non-standard expansion history prior to recombination which our formalism does not take into account. This means it is possible that some other regions of parameter space that are not constrained by the suppression of neutrino freestreaming are in fact constrained by other effects (such as an enhanced Silk damping) which we do not account for here. Accounting for this non-standard expansion history, however, requires a dedicated analysis that is beyond the scope of this work.

\textit{--- $\nu_i \leftrightarrow \nu_j \,\phi$: Neutrino decays.} The fact that neutrino decays could impact neutrino freestreaming was pointed out almost 20 years ago~\cite{Hannestad:2005ex}. Since then, the impact of neutrino decays on CMB observations has been studied by many authors~\cite{Basboll:2008fx,Escudero:2019gfk,Chacko:2019nej,Chacko:2020hmh,Barenboim:2020vrr,Chen:2022idm,Abellan:2021rfq}. In this context, very recently, Refs.~\cite{Barenboim:2020vrr,Chen:2022idm} have made a significant step forward by providing the first calculation of the actual collision term for neutrino decays in the early Universe. This supersedes previous analyses using heuristic arguments to model the rate at which neutrino freestreaming is suppressed~\cite{Hannestad:2005ex,Archidiacono:2013dua,Escudero:2019gfk}. In particular, Ref.~\cite{Barenboim:2020vrr,Chen:2022idm} now find that the actual rate suppressing neutrino freestreaming is $\Gamma_{\rm nfs} \propto \Gamma_\nu \,(m_\nu/T)^5(\Delta m_\nu^2/m_\nu^2)^2$ where $\Gamma_\nu=1/\tau_\nu$ is the neutrino decay rate in vacuum and in the neutrino rest frame, and $\Delta m_\nu^2$ the mass-squared splitting between the heavier and lighter neutrino mass states. By doing a Planck legacy analysis, Ref.~\cite{Chen:2022idm} finds constraints on the neutrino lifetime that depend upon the specific decay channel but that can reach $\tau_\nu \gtrsim 10^7\,{\rm s}$ at 95\% C.L. In our case, for neutrinos with masses $m_\nu \lesssim 0.1\,{\rm eV}$, the power-law rate with $n_{\rm int} = -5$ matches the behavior of the actual rate which is explicitly described in Eq.~(13) of~\cite{Chen:2022idm}. Translating our bound for this rate of $z_{\rm int}< 1300$ at 95\% C.L.\ into a neutrino lifetime requires specifying the number of neutrinos decaying and also the relevant decay channels which in turn fixes the rate as the neutrino mass differences are known. For the purpose of illustration, consider the case of inverted ordering, with two neutrinos decaying at the same rate, and assuming the lightest neutrino to be massless. In this case $\nu_1$ and $\nu_2$ are almost degenerate and $\Gamma_{\rm nfs} \simeq (0.05/\tau_\nu)\,(m_\nu/T_\nu)^5$. Using our bound of $z_{\rm int}< 1300$ one can find a constraint on the neutrino lifetime of $\tau_{\nu_{1,2}} > 4 \times 10^8\,{\rm s}$ at 95\% C.L. This should be compared to the result of Ref.~\cite{Chen:2022idm} which for this scenario has explicitly found $\tau_{\nu_{1,2}} > 5.5 \times 10^7\,{\rm s}$ at 95\% C.L., which within our parametrization would correspond to $z_{\rm int} < 1800$. These numbers are comparable and we attribute the difference to two factors. Firstly, we are running an analysis over $\log_{10}(z_{\rm int})$ within a rather wide region, while Ref.~\cite{Chen:2022idm} uses a linear prior for the neutrino freestreaming rate which in turn covers a much more restrictive redshift range. This means that parameter space volumes can affect to some degree the bound on $z_{\rm int}$. Secondly, given that the bound on $\tau_\nu$ scales approximately as $(1+z_{\rm int})^7$ a small variation at the analysis level can easily account for the difference in the lifetime, and indeed $(1800/1300)^7\simeq 10$.

\textit{A particle physics model for $n_{\rm int} = 3$?} Our analysis of Section~\ref{sec:results_models} has shown that neutrinos interacting with a rate $\Gamma_{\rm nfs} \propto T^3$ (i.e.\ with $n_{\rm int} = 3$) could interact until much lower temperatures than those allowed for the cases with $n_{\rm int} = 4,\,5$, while being in agreement with Planck CMB data. This happens because the redshift dependence of the rate is comparable to $H(z)$ and the suppression of neutrino freestreaming can then be compensated at the CMB level with shifts in $n_s$, $A_s$ and $H_0$. So far, however, there is no particle physics model in the literature known to us that possesses this rate. Nevertheless, it may not be complicated to build such a model. For reactions involving two particles $\Gamma = n \left<\sigma v\right>$. Since $n\propto T^3$, this would mean that $\left<\sigma v\right>\sim {\rm constant}$. Of course, the model building difficulty resides in the fact that neutrinos are relativistic at the time of recombination, but it may well be possible to construct a scenario where $\left<\sigma v\right>$ is indeed temperature independent. Although beyond the scope of our study, it would be interesting to consider building such a scenario in light of the non-standard values of  $n_s$ and to a lesser degree $A_s$ and $H_0$ that could be allowed.

In summary, there are several well motivated extensions of the Standard Model where neutrinos can interact and reduce neutrino freestreaming in the early Universe. With this discussion we expect to give particle physicists a guide on how constraints on the neutrino freestreaming window can be translated into particle physics models, such as couplings and masses.

\section{Conclusions}\label{sec:conclusions}

Cosmology has been shown to be a powerful probe of neutrino interactions. In particular, the fact that the CMB is compatible with three freestreaming neutrino species serves as a stringent constraint on many particle physics scenarios, including neutrino self-interactions, neutrino annihilations, eV-scale neutrinophilic bosons, and neutrino decays (see Fig.~\ref{fig:summary} for a summary of the models and the rates they lead to). These scenarios can lead to a suppression of the neutrino anisotropic stress which is particularly constrained by Planck data. 

In this work we have taken a global perspective on this problem with the aim of narrowing down the region of redshifts in which neutrinos have to be freestreaming in order to be compatible with Planck data. In fact, we have seen that this redshift window is somewhat dependent on the precise temperature dependence of the rate that suppresses neutrino freestreaming as can be seen from Tab.~\ref{tab:cosmo}. Nevertheless, these results globally show that neutrinos should freestream at $2000 \lesssim z\lesssim 10^5$ in order to be in agreement with Planck data. The only exception to this is the case of neutrinos interacting with $\Gamma_{\rm nfs}\propto T^3$, where the temperature dependence of $\Gamma_{\rm nfs}$ and $H$ is very similar. In this case, we have found a degeneracy with $n_s$ (and to a certain extent $A_s$ and $H_0$) that allows for significant interactions even around the epoch of recombination.

Importantly, we have also considered interaction rates that are transient. This captures scenarios in which neutrinos are freestreaming at some high redshift, then they become interacting, and eventually as the Universe cools down they become freestreaming again. Our results for this type of scenario are most clearly summarized in Fig.~\ref{fig:envelopes}. This, together with the results on neutrinos interacting via rates that are power-laws in temperature, allow us not only to corner the redshift window where neutrinos need to freestream, but also to bound how large the neutrino interaction rate can be within these redshifts. In particular, we have found that across this redshift window $\Gamma_{\rm nfs}/H\lesssim 1-10$, while outside of the window a substantially larger ratio is allowed. 

In this work we have taken a simplified and model independent approach to account for the suppression of the neutrino anisotropic stress in the early Universe. However, many of the rates that we consider can actually be mapped to relevant particle physics scenarios. In particular, we discussed such a mapping in Section~\ref{sec:applications}. Although practitioners should use the rates and results for models that are already in the market from the dedicated analyses in the literature, we hope that our discussion can be useful to guide model builders interested in estimating bounds within scenarios beyond the Standard Model that are not yet discussed, without the need to perform a dedicated Planck legacy analysis.

Finally, in Section~\ref{sec:CMBS4} and Section~\ref{sec:LSS} we have discussed how future data can improve Planck's legacy by expanding the extent and depth of the neutrino freestreaming window. In particular,  we have seen that CMB-S4 could not only expand the upper limit of this window from $z_{\rm int} \simeq 10^5$ to $z_{\rm int}\simeq 3\times 10^5$, but it is also expected to be sensitive to transient rates suppressing neutrino freestreaming that can be up to an order of magnitude smaller than those currently excluded by Planck, see Fig.~\ref{fig:CMBS4forecast}. Moreover, we find that the degeneracy of the neutrino interaction strength with $n_s$ allowing for large interactions around recombination for the special case  $\Gamma_\text{nfs}\propto T^3$ can be broken by Stage-IV CMB experiments. In addition, we have explored the implications of interacting neutrinos for galaxy surveys. We have seen that the effect of neutrinos interacting at low redshift can only lead to a small effect on the matter power spectrum which is not expected to be observable given Planck constraints. On the other hand, we found that it is in principle possible for galaxy surveys such as DESI to still probe scenarios where the neutrinos interact only at high redshift, see Fig.~\ref{fig:pk_frac_diff}. For these scenarios, we have in addition calculated the 1-loop correction in order to highlight where non-linear effects can be relevant for future analyses.

In conclusion, we have investigated the model dependence of CMB constraints on neutrino interactions, and established the existence of a \emph{freestreaming window} using current Planck and BAO data. We find that the freestreaming window is model-independent under some broad assumptions, with a notable exception of neutrino interactions for which the interaction rate evolves similarly in time as the Hubble rate, $\Gamma_{\rm nfs}/H \sim T$. We have calculated how future CMB and galaxy surveys can improve upon Planck's legacy, and furthermore break a degeneracy with $n_s$ (and $A_s$ as well as $H_0$) that exists for the special scenario of slowly time evolving ratio $\Gamma_\text{nfs}/H$. These results are relevant for a variety of scenarios beyond the Standard Model where neutrinos interact. Our results can be useful for determining whether a given model is excluded by current cosmological data or whether it could be tested by future surveys.

\begin{center}\textbf{Acknowledgments}  \end{center}

ME is supported by a Fellowship of the Alexander von Humboldt Foundation. MG and PT are supported by the DFG Collaborative Research Institution Neutrinos and Dark Matter in Astro- and Particle Physics (SFB 1258).

We acknowledge the use of the python package GetDist~\cite{Lewis:2019xzd} for posterior plots.
\vspace{0.2cm}

\appendix
\section{Approximations}\label{sec:approximations}

Solving the full Boltzmann hierarchy~\eqref{eq:Boltzmannhierarchy}%
\footnote{We truncate the Boltzmann hierarchy in the numerical solution at
$l_{\mathrm{max}} = 17$ (CLASS default).}
is computationally very expensive, and for MCMC exploration it is preferable
to use a more efficient method to integrate the equations. In this appendix we discuss the
approximation schemes we utilize. Our overall strategy is to use the default
CLASS fluid approximation for non-cold relics (CLASS-FA from here on) whenever
appropriate, which has been shown to work at the sub-permille level in
$\Lambda$CDM for phenomenologically relevant neutrino
masses~\cite{Lesgourgues:2011rh}. In addition, for cases with three interacting
neutrinos we assume that the neutrinos are degenerate and solve the hierarchy
for one neutrino species.

Depending on the interaction type, we use the following approximation schemes:
\begin{itemize}
    \item \emph{Power-law, $\nint = [3, 4, 5]$}: The interaction is efficient
        at high redshift, therefore for a given mode $k$, we use the full
        Boltzmann hierarchy up until the point where $k \tau > 31$%
        \footnote{This corresponds to the default settings of CLASS.}
        at which CLASS-FA is turned on.
        For the $95\%$ C.L.\ $\zint = 85000$ for $\nint=5$ this means that
        the fluid approximation is turned on for wavenumbers $k \gtrsim
        9~h/\mathrm{Mpc}$ when $z=\zint$. We compare the fluid approximation to
        using the full hierarchy at all times for a grid of $\zint$ parameters
        in the prior range (see Eq.~\eqref{eq:nint_prior}) and find differences on
        the $C_l$'s that are negligible compared to cosmic variance.
    \item \emph{Power-law, $\nint = [-5,-3,-1,1]$}: The interaction is efficient
        at low redshift, therefore we use a modified CLASS-FA where we set
        explicitly the velocity dispersion $\sigma$ to zero in the fluid
        equations. This modified approximation is turned on when
        $k\tau > (1/\mathrm{Mpc})\,\tau_{\mathrm{int}}$. In order words, all
        modes $k > 1~\mathrm{Mpc}^{-1}$ are treated by the modified CLASS-FA
        when $z = \zint$. We checked that the error from the fluid
        approximation is negligible compared to cosmic variance for a grid
        of $\zint$ parameters in the prior range.
    \item \emph{Transient interaction, $b=[0,2,4]$}: The interaction is efficient at an
        intermediate redshift, so we utilize CLASS-FA, making sure that it is
        only turned on when the interaction is small compared to the Hubble
        rate. More specifically, given $z_{\mathrm{int}}^{\mathrm{max}}$ and
        $\Gamma/H\big|_{\mathrm{nfs}}^{\mathrm{max}}$ we compute the time
        $\tau'$ at which $\Gamma_{\mathrm{nfs}}/H < 10^{-4}$ (after the
        interaction has reached its maximum), and allow the solver to switch to
        CLASS-FA when both $k\tau > 31$ and $\tau > \tau'$ are satisfied.
        Comparing this approximation scheme to the full hierarchy, we find
        negligible differences in the $C_l$'s for an evenly spaced grid of 
        $\mathcal{O}(100)$ points in 
        $\log_{10} z_{\mathrm{int}}^{\mathrm{max}}$--%
        $\log_{10} \Gamma/H\big|_{\mathrm{nfs}}^{\mathrm{max}}$ parameter space.
\end{itemize}
For the runs with $N_{\mathrm{int}} = 1$ or $N_{\mathrm{int}}=2$ interacting neutrinos, we do not
use any approximation scheme and always use the full Boltzmann hierarchy for the neutrinos.

Finally, we checked the initial conditions for the neutrino perturbations in the case where the interaction is efficient at early times.
In particular, we found negligible differences in the angular power spectra when imposing the initial condition $\Psi_{l\geq 2}(\tau_{\mathrm{ini}}) = 0$ compared to the standard initial condition~\cite{Ma:1995ey}.

\begin{figure*}[t]
    \centering
    \includegraphics[width=0.9\textwidth]{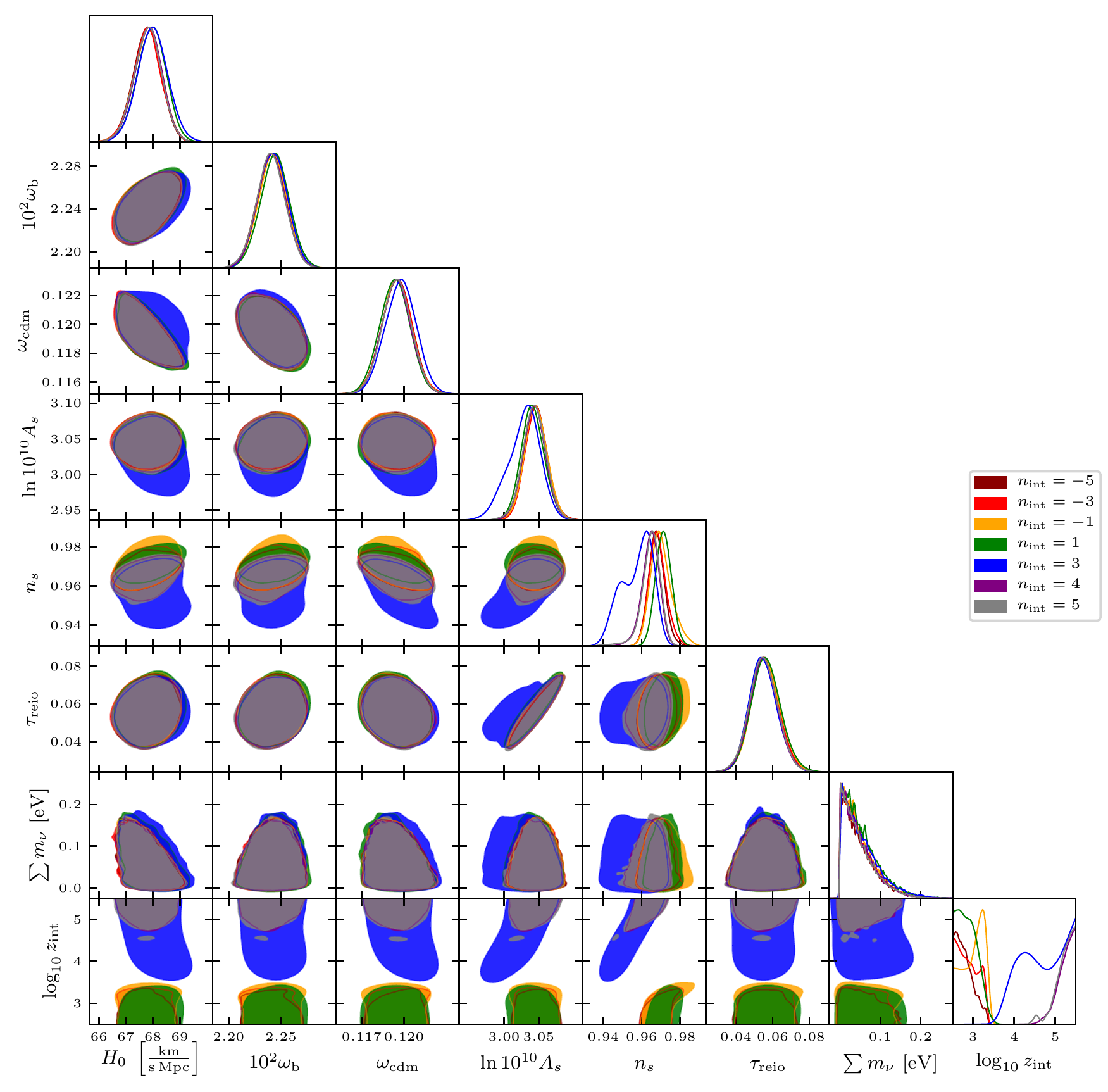}
    \caption{Full parameter constrains for the power-law interaction scenarios, corresponding to Fig.~\ref{fig:posterior_nint_all}. We note that for $n_{\rm int}\not=3$ there are no strong correlations between $z_{\rm int}$ and any cosmological parameter apart from the spectral index $n_s$. It can clearly be seen that the $n_{\rm int} =3$ case is special and we display it in more detail in Fig.~\ref{fig:posterior_nint_3}.}
    \label{fig:posterior_nint_full}
\end{figure*}

\section{Additional results for the $n_{\rm int}= 3$ case}\label{sec:nint3appendix}

\begin{figure*}[t]
    \centering
    \includegraphics[width=0.9\textwidth]{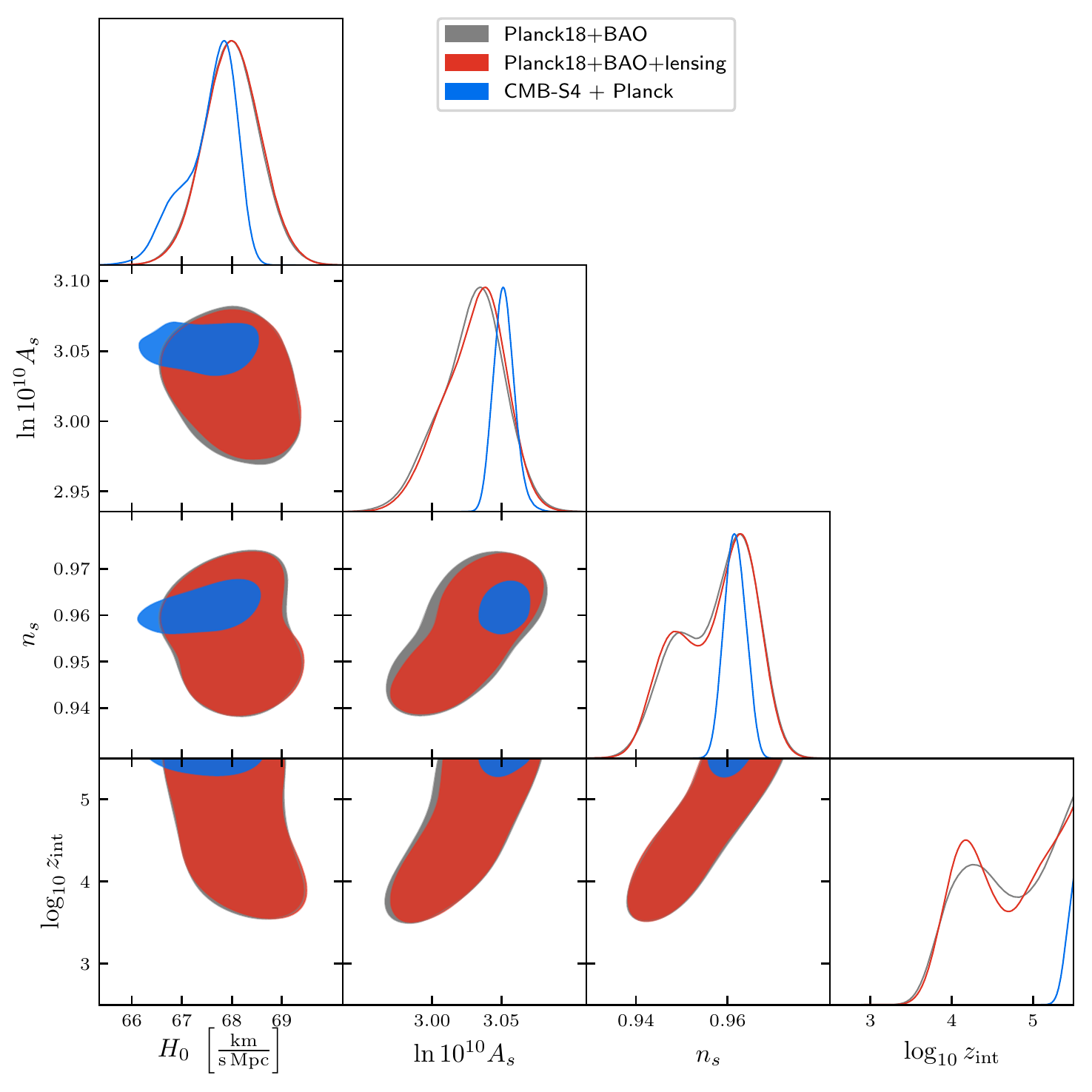}
    \caption{Posterior probability contours for the case of neutrinos interacting with a power-law with $n_{\rm int} = 3$. We show $H_0$, $A_s$, $n_s$ and $z_{\rm int}$. We display in grey Planck+BAO, in red Planck+lensing+BAO, and in blue the forecast of CMB-S4 assuming the true model to be $\Lambda$CDM 
    with $\omega_{\mathrm{b}} = 0.02239$, $\omega_{\mathrm{cdm}} = 0.1199$, $H_0 = 68$\,km/s/Mpc, 
    $\ln 10^{10} A_s = 3.0521$, $n_s = 0.9659$, $\tau_{\mathrm{reio}} = 5.7367\times 10^{-2}$
    and $\sum m_{\nu} \simeq 0$.}
    \label{fig:posterior_nint_3}
\end{figure*}

As discussed in the main text neutrinos can interact at much lower redshifts than naively expected for a power-law interaction rate with $n_{\rm int} =3 $. This happens because in this scenario the time dependence of the rate is very similar to the time dependence of the Hubble parameter, which means that neutrinos can interact over a large window of redshifts. This in turn allows the effect on the angular power spectra to be approximately compensated by variations of standard cosmological parameters. In particular, $n_s$ and to a lesser extent $A_s$ and $H_0$. Since this affects $n_s$ and $A_s$ one should wonder whether Planck lensing data could reduce the degeneracy. In this context we run another analysis including this data set and we show it in red in Fig.~\ref{fig:posterior_nint_3}. We see that actually the ability of lensing to reduce this degeneracy is very modest. In addition, we also  investigate how the CMB-S4 experiment can test this model. The results are shown in blue. We can clearly appreciate that CMB-S4 will be capable of breaking the degeneracy and reaches a 95\% C.L.\ sensitivity of $z_{\rm int} < 2.4\times 10^5$. We expect that the improved precision of the polarization spectra at $\ell \lesssim 1000$ as compared to Planck plays an important role, along with improvement of both temperature and polarization data up to $\ell \sim 3000$.

\newpage

\bibliography{biblio.bib}

\end{document}